\documentclass[notitlepage]{revtex4-1}
\usepackage{graphicx}
\usepackage{subfigure}
\usepackage{relsize}
\usepackage{amsmath}
\usepackage{amsfonts}
\usepackage{amssymb}
\usepackage[dvipsnames]{xcolor}
\usepackage{color} 
\usepackage{tikz}
\usetikzlibrary{shapes}
\usepackage{float}
\usepackage{tabularx}

\definecolor{mypink3}{rgb}{0.858, 0.188, 0.478}

\begin{document}
\title{Experimental studies on the frequency selection in flat plate wakes: \\mean flow stability analyses and low dimensional modeling}
\author{Dipankar Dutta$^{1}$, Indra Kanshana$^{1}$, Shyam Sunder Gopalakrishnan$^{2, 3,\dagger}$, and A. C. Mandal$^{1,\ddagger}$}

\affiliation{$^1$ Department of Aerospace Engineering, Indian Institute of Technology, Kanpur 208016, India \\
$^2$ Facult\'{e} des Sciences, Universit\'{e} libre de Bruxelles (ULB), CP. 231, 1050 Brussels, Belgium \\
$^3$ Laboratoire de Physique des Lasers, Atomes et Mol\'{e}cules, CNRS UMR 8523, Universit\'{e} Lille 1 - 59655 Villeneuve d'Ascq Cedex, France}

\email{$\dagger$ shyam7sunder@gmail.com \\
$\ddagger$ alakeshm@iitk.ac.in }

\date{\today}
\begin{abstract}

We investigate the global frequency selection of two-dimensional vortex shedding in the flat plate wake. The analysis is based on the mean flow velocity profiles obtained from experimental measurements carried out for two values of Reynolds number, 1850 and 3350, which are based on the plate thickness and the free-stream velocity. Two different trailing edge geometries of the flat plate are considered in this study: blunt and circular. By performing local spatio-temporal analyses on the measured mean flow velocity profiles, we estimate the global shedding frequency of the flow. This is in excellent agreement with the shedding frequency measured experimentally. To complement the study, we carry out a low-dimensional modeling based on the proper orthogonal decomposition (POD) of the flow fields which is novel for flat plate wakes. We observe that a model based on only two POD modes produces an accurate estimate of the global shedding frequency. Our results also highlight the role of the nonlinear interaction strength between the mean flow with the higher harmonics thereby experimentally supporting the theoretical criterion outlined in \cite{sipp2007global}.
 \end{abstract}
 \maketitle

\section{Introduction}

Ever since the seminal work by Strouhal \cite{Strouhal1} the frequency selection of periodic vortex shedding in the wakes of flows past obstacles have been of interest to both researchers and engineers \cite{Roshko1,bauer1961vortex, Bloor1, Huerre1, Will2}. 
With the advances in experimental measurements, the frequency selection criterion has been proposed \cite{Chomaz1, Will1, Will2}  and  has been successfully built in the framework of weakly non-parallel flows \cite{Monk1, Huerre1, Monk2, Chomaz1, Pier1, Pier2}.  Interestingly, the selection criterion when applied to the mean flow of bluff-body wakes have shown to yield particularly good results for the prediction of the frequency of the unsteadiness. This has been observed in the wake formed by a rectangular forebody \cite{Hammond1}, and by a cylinder \cite{Pier2}, and has been confirmed by performing a global stability analysis on the mean flow of a cylinder wake \cite{barkley2006linear,mittal2008global}. The theoretical support for these observations were advanced by \cite{sipp2007global} where they outlined the conditions when the stability of a mean flow would approximately match with the nonlinear frequency of the unsteady wake. Using a weakly global nonlinear analysis in the vicinity of the critical Reynolds number, they showed that the frequency selection is governed by the relative interaction strength of the mean flow and the higher harmonics. Specifically, the global frequency is well-predicted by a mean flow analysis when the nonlinear interaction of the zeroth (mean flow) harmonic with the first harmonic is way higher than the interaction of the second harmonic with the first harmonic.
They observed that the linear stability analysis of the mean flow yields good results for a cylinder wake, but does not provide a meaningful insight in the case of an open cavity flow. In addition, it has been noted in \cite{Pier2, Hwang1,Luchini1} that the global frequency is well estimated by the local absolute frequency at the stagnation point in the flow.
In this study, we address these concerns for flat plate wakes using experimentally measured mean flow velocity profiles which is missing in the literature.

The wake formed by the trailing edge of a flat plate, or a thin aerofoil at zero incidence, has been extensively studied in the past in the laminar, transitional, and turbulent regimes \cite{Sato61,koku70,nish78,rama82,wygn86,jovi86,bogu88,juli03,lash90,nara72}. Understanding the frequency selection in flat plate wakes is of particular importance as they find direct applications in the aerospace industry, where drag reduction and noise control are desired \cite{jovi86,cari17}, in the design of turbine and compressor blades \cite{bogu88}, and from an academic point of view. The flat plate wake is influenced by the profile of the trailing edge. Carefully made measurements by \cite{nish78} has showed that the wake behind a flat plate with a sharp trailing edge can be kept laminar even at a Reynolds number of 3000 (based on the freestream velocity and the plate length), even though a wake flow is inherently unstable. The turbulent wake characteristics of flat plates with a sharp trailing edge have been well investigated \cite{bogucz1988turbulent,rama82,jovi86}. In contrast, there exists vigorous vortex shedding for flat plates with a blunt trailing edge \cite{nakamura1991experiments,rai16}. 
The vortex shedding from flat plates with square leading and trailing edges were experimentally studied in \cite{nakamura1991experiments} where they showed that the separated shear layer becomes unstable downstream of the trailing edge corner. As maintaining extremely sharp trailing edges for flat plates still pose as engineering challenges, it is pertinent to study the effect of the trailing edge geometry, and is thus considered in the present study. In addition to the trailing edge geometry \cite{Siev90,Tayl11,rai16}, the characteristics of the flat plate wake depends on the velocities in the upper and lower surfaces of the plate \citep{Bold76}, and various other boundary layer parameters such as the boundary layer thickness and shape factor \cite{Rowe01,Durg13,Siev90}.

One of the first experimental measurements on the vortex shedding behind thin flat plates were made by \cite{bauer1961vortex}. The hydrodynamic resonance criteria advanced by \cite{koch1985local} based on a local stability analysis predicts well the observed shedding frequency in \cite{bauer1961vortex}, which is governed by the streamwise location where the instability character changes from an absolute to a convective nature. Apart from the global frequency selection criterion proposed in \cite{koch1985local}, there are  other criteria such as the maximum growth criterion by \cite{pierrehumbert1984local}, and the initial growth criterion by \cite{monkewitz1987absolute}.
In a strictly linear setting, the first rigorous selection criterion has been established in \cite{Chomaz1}, with the global frequency given by a saddle-point condition based on an analytic continuation of absolute instability frequency in the complex $x = (x_r,x_i)$ plane. This criterion is seen to accurately predict the selected frequency in the wake of a blunt-edged plate \cite{Hammond1} using the time-mean flow. Since then the non-linear instability of slowly divergent flows has been addressed in \cite{Pier1,Pier2} with the bifurcation to self-sustained oscillation in two-dimensional wake flows being triggered whenever a region of local absolute instability exists in the flow. The dominant shedding frequency is imposed by the first absolutely unstable downstream station. 
As observed in \cite{Hammond1}, it has been shown in \cite{Pier2} that the time average of the oscillating wake behind a cylinder provided the best profile to predict the vortex shedding frequency. 

It is to be noted that the cylinder wake studied in \cite{Pier2}  goes from a locally convective nature to a finite sized absolutely unstable region, with the frequency at this transition station fixing the global shedding frequency, arising from a balance between 
upstream perturbation growth and downstream advection \cite{dee1983propagating,Pier2}. 
However, the wake of the blunt-edged plate studied in \cite{Hammond1} is locally absolutely unstable at the blunt trailing edge. 
The first absolutely unstable downstream station in this case would correspond to the streamwise location proposed in \cite{koch1985local}. This observation is detailed in the criterion proposed in \cite{monkewitz1987absolute} where the initial growth of the disturbances within the convectively unstable region downstream of the bluff-body, the extent of which then subsequently fixes the spatial scale associated with the global shedding frequency.
As mentioned earlier, using a weakly global analysis \cite{sipp2007global} theoretically outlined the conditions when the frequency of the time-averaged mean flows 
matches with the global shedding frequency of the wake. This is governed by the interaction strength of the higher harmonics with the mean flow harmonic. These observations have been extended and verified for the wake of a cylinder using time-averaged mean flows in \cite{khor2008global}. Such an analysis is missing in the literature for flat plate wakes which forms the objective of the present study.
We investigate the above discussed characteristics for the wake of a flat plate with a blunt/circular trailing edge by performing a local stability analysis using the time-averaged flows, and compare them with the global shedding frequency. In addition, we also aim to support the theoretical criterion presented in \cite{sipp2007global} using the experimentally obtained flow fields.

The characteristics of the wake formed by a circular cylinder, and by a flat plate with a circular trailing edge are indeed different. This aspect is studied in \cite{Ryan05} where they carried out a Floquet stability analysis to investigate the transition scenario in the wake behind a flat plate with a circular trailing edge, referred to as an elongated cylinder with an aerodynamic leading edge, and with a blunt trailing edge. Their study revealed the existence of three different modes, showing similarities and differences between the two different wakes, concluding that the transition scenario in the wake of a flat plate with a circular trailing edge may not be completely generic, as often assumed based on cylinder wakes. 
The observations made in \cite{Ryan05} have been experimentally confirmed in \cite{Nagh12,Nagh14,Dodd10} wherein they indicate that the wake topology and their instability at the trailing edge of a flat plate are different from those observed in the case of a compact body. Furthermore the wake characteristics of a flat plate with circular and elliptic trailing edges were numerically studied in \cite{rai16}, where they concluded that the streamlining of a trailing edge results in weaker vortex shedding, with a smaller separated region.

Similarly, various studies have revealed the differences in the secondary instability mechanisms and the resulting vortical structures downstream of a flat plate \cite{julien_lasheras_chomaz_2003,julien2004secondary} from that of a cylinder wake \cite{leweke1998three,barkley2006linear}. The flat plate wake flows were experimentally studied at low Reynolds numbers in \cite{julien_lasheras_chomaz_2003} which were formed by the merging of two parallel laminar streams at the trailing edge.
It is observed that the leading modes are characterised by the same wavelength unlike in a cylinder wake, indicating that they arise from the same mechanism, which is a combination of hyperbolic and elliptical instability. These observations have been corroborated numerically in \cite{julien2004secondary}. The differences in bluff body wakes  from that of a flat plate stem from the fact that the shear layers in bluff-body wakes separate at a certain location on the surface of the body, thus introducing a second characteristic length in the problem. Whereas in a flat plate wake, the flow separation occurs right at the trailing edge.
 In addition, the base flow varies rapidly behind a cylinder and is characterised by flow reversal with highly concentrated vorticity regions. In this region, the secondary instability is absolutely unstable  due to a global secondary bifurcation \cite{barkley1996three} whereas for a flat plate the instability is convective \cite{julien_lasheras_chomaz_2003}.
 
 \begin{figure}
 \unitlength=80.0mm
\centerline{
\includegraphics[width=1.15\unitlength,height=0.95\unitlength]{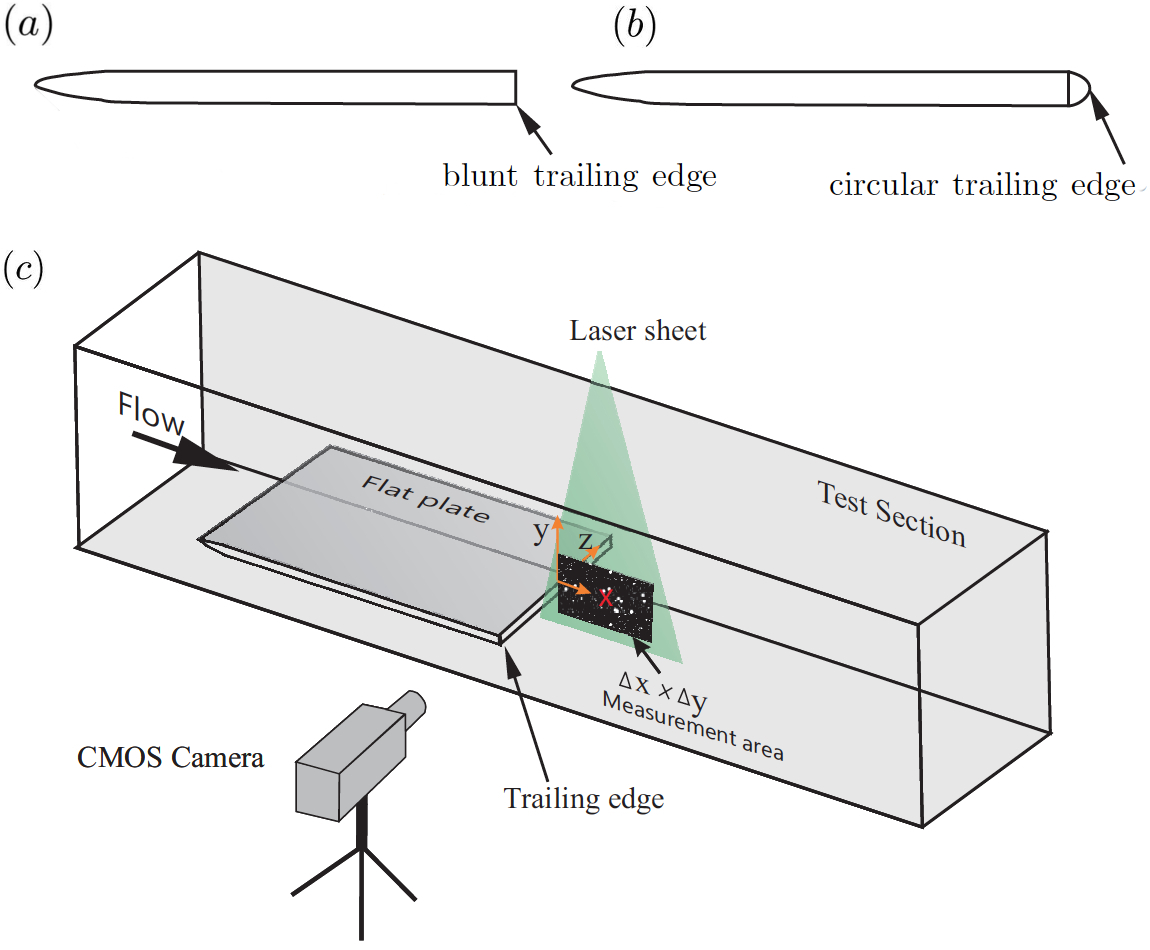}
}
%\begin{picture}(0,0)
%\put(-0.75,0.0){(a)}
%\put(0.65,0.0){(b)}
%\end{picture}
 \caption{The experimental set-up used in the present study on flat plate wakes. Two profiles of the flat plate are considered, with a completely blunt $(a)$ and circular $(b)$, trailing edges respectively. $(c)$ Illustration of the TR-PIV measurement setup.}
 \label{fig:schem}
 \end{figure}

In this work, we investigate the global frequency of flat plate wakes by performing a local stability analysis on the mean flow fields. A blunt (square/rectangular) and a circular profile for the trailing edge geometry are considered in this study which are shown in Fig. \ref{fig:schem}(a,b). The experiments were conducted in a wind tunnel, a schematic sketch of which is shown in Fig. \ref{fig:schem}(c). The two-dimensional velocity fields were obtained by using a Time Resolved Particle Image Velocimetry (TR-PIV) technique. The study is carried out for two values of the Reynolds number, 1850 and 3350, based on the freestream velocity and the plate thickness. Using the obtained flow fields, we perform a local spatio-temporal stability analysis on the mean velocity profiles at different streamwise locations of the unsteady wake. The selected global frequency is then compared with the different characteristic frequencies discussed earlier. It has been observed that the local absolute frequency at the end of the absolutely unstable domain of the flow predicts the global shedding frequency. A observed in other bluff-body flows, the global frequency of the wake is well predicted by the local absolute frequency at the stagnation point in the flow.  An excellent pedagogical review on the theoretical tools used in the present study can be found in \cite{Huerre2}.

%As observed in earlier studies on other bluff-body flows, the global frequency of the wake is well predicted by the local absolute frequency at the stagnation point in the flow.  An excellent pedagogical review on the theoretical tools used in the present study can be found in \cite{Huerre2}

To complement the study, we analyse the experimental data by carrying out a proper orthogonal decomposition (POD) of the wake velocity fields. POD is a useful tool, especially in experimental investigations, which is capable of extracting information using snapshots of the flow fields  \cite{aubr88,noac03,berk93,remp94,siegel_seidel_fagley_luchtenburg_cohen_mclaughlin_2008,gall04,luch09,Mandal1}. 
It provides a set of orthogonal and optimal basis functions, onto which the Navier--Stokes equations can be projected to construct a reduced order model \cite{remp93,raja94}. We perform the POD analysis on the fluctuations of the wake velocity fields by removing the mean flow. It is observed that over 80$\%$ of the energy of the unsteady wake is captured by the first ten POD modes and over 70$\%$ of the energy by the first two POD modes, even for higher Reynolds number considered here, indicating a low dimensional nature of the flow under consideration. Therefore, a low dimensional model based on the dominant POD modes is found to yield an accurate estimate of the selected global frequency. Similar observations have been reported for a cylinder wake using POD analyses \cite{siegel_seidel_fagley_luchtenburg_cohen_mclaughlin_2008}. In doing so, we experimentally support the theoretical criterion outlined in \cite{sipp2007global} which determines whether a mean flow stability analysis is meaningful and can predict the global shedding frequency, based on the nonlinear interaction strength of the mean flow with its higher harmonics.

To this end, we organise the paper as follows. The flow configuration, along with the theoretical, experimental, and numerical tools used in the current study are briefly outlined in section II. Following this, in section III we present the mean flow velocity profiles using which the spatio-temporal stability analyses are performed. The results from POD analysis with a description of the flow features of a flat plate wake are presented in section IV. The paper finishes with a discussion on the global shedding frequency of flat plate wakes along with some concluding remarks in section V.

\section{Flow configuration and methodology}

\subsection{Experimental set-up}

\begin{figure}[H]
 \unitlength=50.0mm
\centerline{
\includegraphics[width=1.967\unitlength,height=1.112\unitlength]{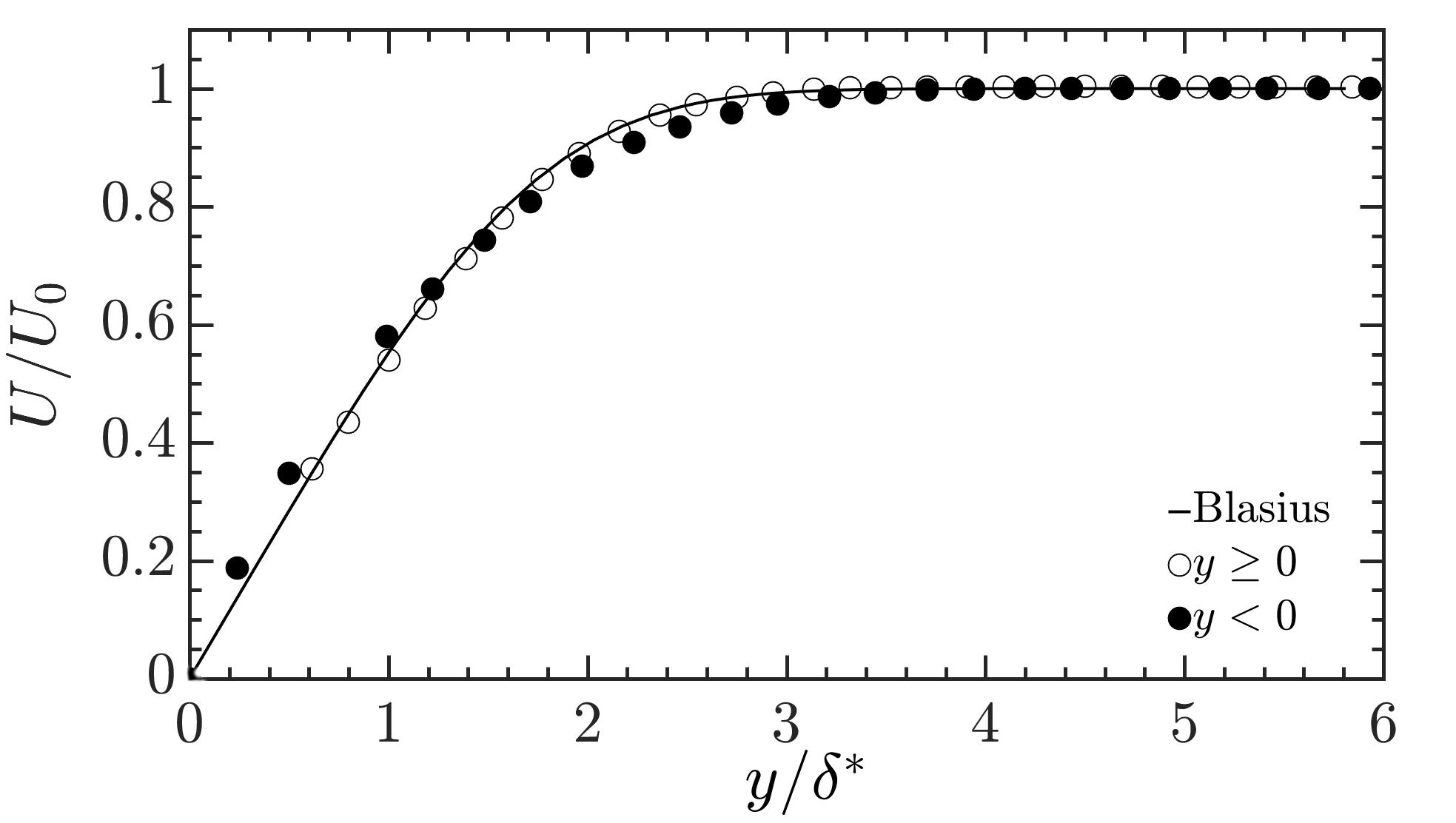}
}
 \caption{Boundary layer profile at $.45D$ (i.e. at $x= - 0.45D$) from the blunt trailing edge of the flat plate for $Re = 3350$. Open and filled circles denote the velocity profiles above and below the flat plate, respectively. The Blasius solution is shown using the solid black line.
}
 \label{fig:blasius}
 \end{figure}

The general flow configuration consists of an incompressible fluid stream over a flat plate with a blunt (circular) trailing edge, a schematic of which is shown in Fig. \ref{fig:schem}. Shape of the leading edge of the plate is a super ellipse, details of which are available in~\cite{Bala17}. A wake forms downstream of the flat plate from the fixed separation points at the trailing edge. Whole field measurements in the wake of the flat plate were carried out using the time-resolved particle image velocimetry (TR-PIV) technique in a low-speed wind tunnel. The wind tunnel used in the present study is an open circuit suction type wind tunnel. The settling chamber of this tunnel houses a honeycomb and six turbulence reduction screens. However, the settling chamber is followed by a contraction cone of $16:1$. The $3000$ mm long test section of the tunnel has a square cross-section of $610$ mm $\times$ $610$ mm, and is followed by a diffuser. The three bladed fan of this tunnel is powered by a $14.5$ kW alternating current motor. The tunnel turbulent intensity measured in the test section is found to be $0.1\%$ of the freestream velocity (see for details \cite{Bala17}). The flat plate has a thickness of $12$ mm and a length of $700$ mm, and is mounted horizontally in the mid-plane of the test section of the tunnel. In the following $x$, $y$, and $z$ represent streamwise, wall normal, and spanwise directions respectively. The free stream velocity $U_\infty$ and the flat plate thickness $D$ were used as the velocity and length scales, with the Reynolds number given by $Re = U_{\infty}D/\nu$. In the present study two values of $Re$ are considered: $1850$ and $3350$, which are carefully chosen so that the boundary layer remains laminar till it reaches the trailing edge of the flat plate. The boundary layer profile at a distance of $0.45D$ from the blunt trailing edge at $Re = 3350$ is shown in Fig. \ref{fig:blasius}. The velocity profile closely matches with the Blasius solution, thereby ensuring that the boundary layer remains laminar for both values of $Re$ chosen in this investigation.

The TR-PIV measurements were carried out at the trailing edge of the flat plate, as schematically shown in Fig. \ref{fig:schem}. The TR-PIV system consists mainly of a $4$ mega pixels CMOS camera with 365 Hz repetition rate in double exposure mode at full resolution (IDT vision, USA), a Nd-YLF dual PIV laser  with $30$mJ/pulse at $1$ kHz repetition rate (Photonics Industries International Inc, USA), and an IDT synchronizer. Using the inboard memory of the CMOS camera, $365$ TR-PIV image pairs were captured per second at full resolution. The flow was seeded using a fog generator. Using appropriate sheet forming optics, a thin laser sheet of about $1$ mm thickness was produced at the region of interest, as shown in Fig. \ref{fig:schem}. Using the ProVISION XS package procured from IDT, the images of the seeding particles in the streamwise wall-normal ($x$-$y$) plane were then acquired over a region of interest of $\approx$ $66$ mm $\times$ $60$ mm. The acquired images were then processed in ProVISION XS using a correlation window of $32$ pixels $\times$ $32$ pixels. Mean quantities are estimated based on 800 PIV realizations. The ProVISION XS is based on mesh free algorithm as detailed in \cite{Lour00}. This software package and the PIV system were also used in our various previous works [e.g.~\cite{Mand10,Mand11,Phan15,Bala17}].

\subsection{Local stability analysis}

The global shedding frequency of bluff-body wakes can be determined from a local stability analysis based on the 
concept of absolute instability \cite{Chomaz1,Monk1, Monk2,chomaz2005global}. There exists a precise location 
in the complex $x$-plane which acts as a wavemaker for the entire field, thereby fixing the global frequency. Following \cite{Pier1,Pier2,Luchini1}, we boldly ignore the highly non-parallel nature of the flat plate wake, even though the Reynolds numbers are on the higher side of the laminar range. We derive the local characteristics at a streamwise location by freezing the $x$- coordinate, and by performing a linear stability on the measured mean flow velocity profiles, $U(y) = U_b(x,y)$. Linearizing around this basic flow by adding a small amplitude perturbation $(u'_x, u'_y)$, gives us the linearized Navier--Stokes equations

\begin{eqnarray}
\partial_{t}u'_{x} + U\partial_{x}u'_{x} + u'_{y}\partial_{y}U &=&  -\partial_{x}p' + Re^{-1} \nabla^2 u'_x, \nonumber \\
\partial_{t}u'_{y} + U\partial_{y}u'_{y}  &=&  -\partial_{y}p' + Re^{-1} \nabla^2 u'_y, \nonumber \\
\partial_{x}u'_{x} + \partial_{y}u'_{y} &=& 0.
\end{eqnarray}

We look for travelling wave solutions in the form of normal modes as $q'(x,y,t) = q(y)\exp[\dot{\imath}(kx - \omega t)]$ where $\omega (\omega_r + \dot{\imath}\omega_i)$ is the complex frequency, and $k (k_r + \dot{\imath}k_i)$ the complex wavenumber; $q'(x,y,t)=[u'_{x},u'_{y},p']^{T}$. The linear stability of these waves are governed by the Orr--Sommerfeld equation \cite{White06}:

\begin{equation}
[(-\dot{\imath}\omega + \dot{\imath}k U)(D^2 - k^2) - \dot{\imath}k \frac{d^2U}{dy^2} - \frac{1}{Re}(D^2 - k^2)^2]u_{y} = 0.
\end{equation}
\label{eqn:OSS}

 \begin{figure}
 \unitlength=50.0mm
\centerline{
\includegraphics[width=1.212\unitlength,height=1.042\unitlength]{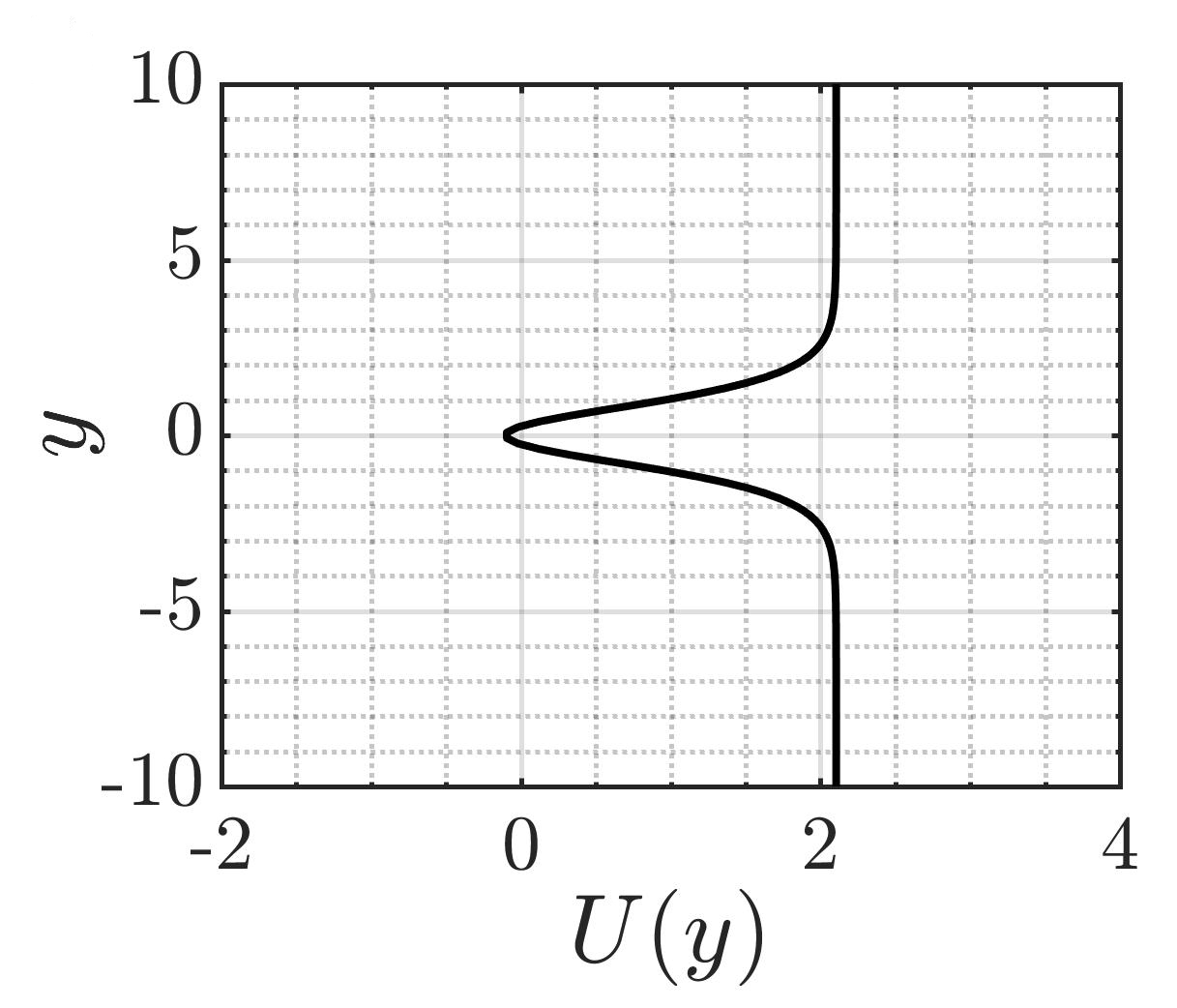}
\includegraphics[width=1.110\unitlength,height=1.095\unitlength]{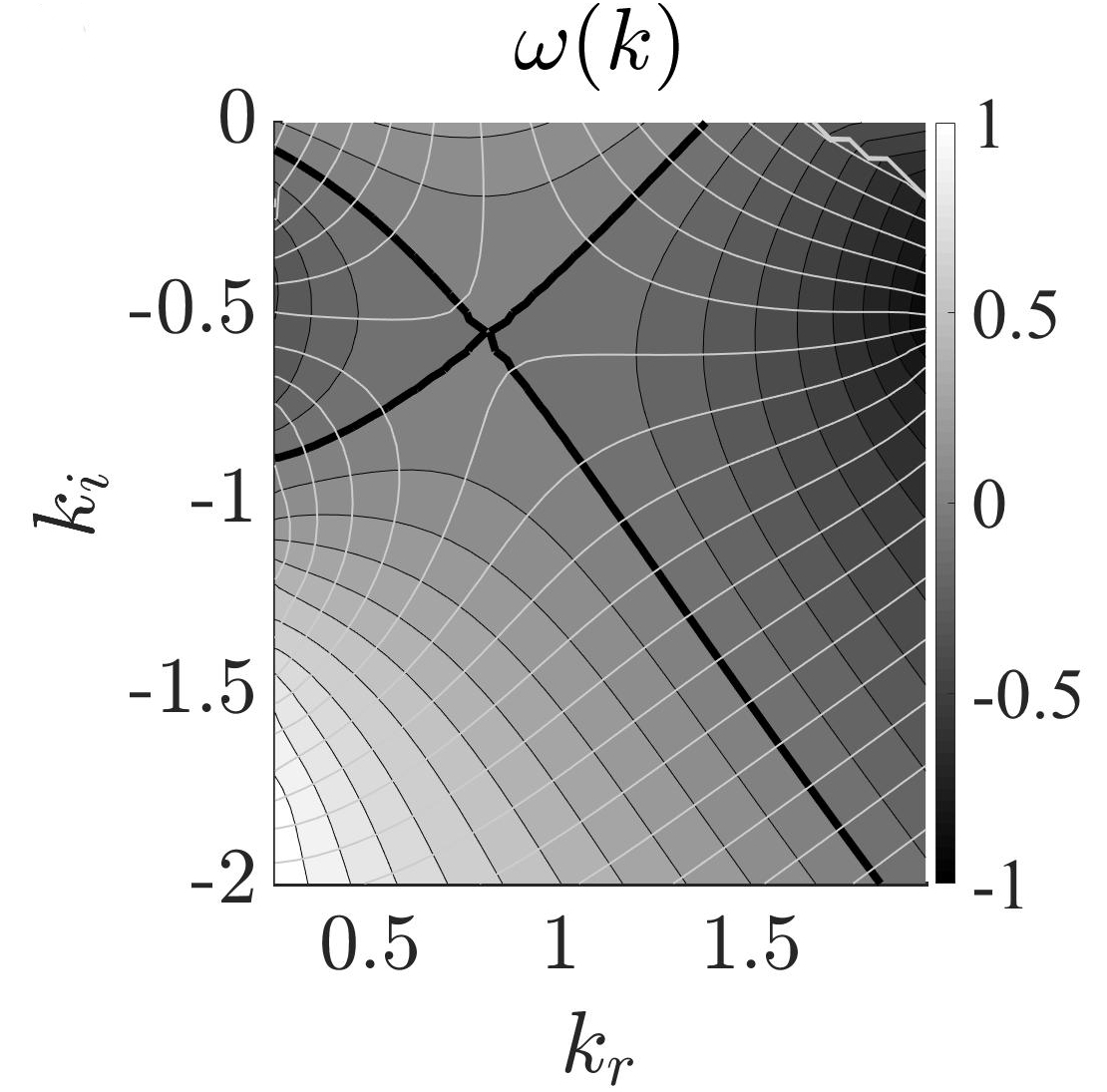}
}
\begin{picture}(0,0)
\put(-1.25,1.05){$(a)$}
\put(0.05,1.05){$(b)$}
\end{picture}
 \caption{(a) Parallel model wake velocity profile $U(y)$ and (b) the contours of $\omega_{\dot{\imath}}$ (greyscale) and $\omega_r$ (greylines) at $Re = 11.3$, $R = -1$, $N =2$.
}
 \label{fig:Monk}
 \end{figure}
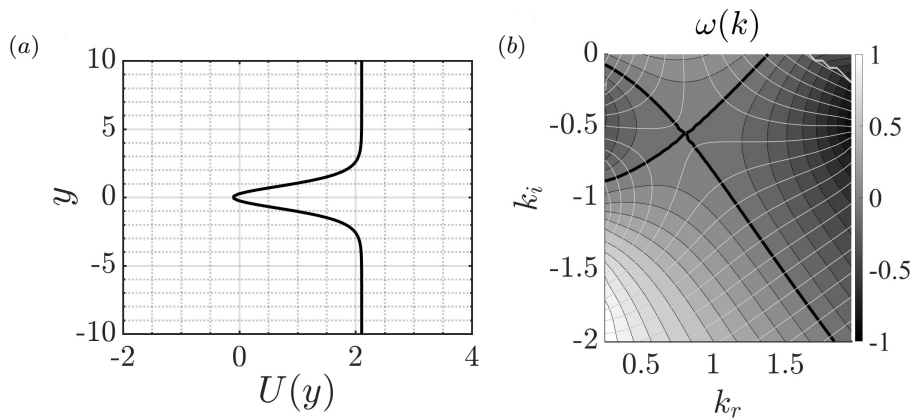 

The Orr--Sommerfeld equation (2) along with the boundary conditions $u_{y} = Du_{y} = 0$ (on the wall and in the free stream) is then solved using standard spectral methods \cite{Trefethen1} which yields the local dispersion relation $\omega = \Omega^{l}(k,x)$. For doing this, the measured mean flow velocity profile at a streamwise location is spatially discretized on $N_y = 100$ Chebyshev collocation points in the wall normal direction ($x_j = \cos\big(\frac{j\pi}{N_y}\big)\epsilon[-1, 1], j = 0, 1, ... N_y$). By applying the classical Briggs--Bers criterion \cite{Bers1, Briggs1, Pier2} of zero group velocity condition the absolute frequency $\omega_0(x)$ is then obtained.
As part of the current study, the spatio-temporal analysis of a parallel model wake flow at a low Reynolds number is carried out. Following \cite{Monk1, Hwang1}, the profile of the basic flow is given by
\begin{eqnarray}
U(y) = 1 - R + 2RF(y), \\
R = \frac{U^{*}_c - U^{*}_{\infty}}{U^{*}_c + U^{*}_{\infty}}, \\
F(y) = [1 + \sinh^{2N}(y\sinh^{-1}(1))]^{-1},
\end{eqnarray}
where $N$ is the shape parameter. The superscript $*$ denotes a dimensional quantity, with $U_c$ being the centerline velocity ($y = 0$), and $U_{\infty}$ the freestream velocity. Fig. \ref{fig:Monk}(a) shows the parallel model wake profile at a Reynolds number $11.3$ (based on the average basic flow velocity and the wake half-width), $R = -1$, and $N = 2$.  The variation of $\omega_r$ and $\omega_{\dot{\imath}}$ in the complex $k$ plane is shown in panel (b). The absolute wavenumber $k_0 = 0.8047 - 0.5569\dot{\imath}$, and the absolute frequency $\omega_0 = 1.0086 + 0\dot{\imath}$ is in excellent agreement with \cite{Monk1}.

\subsection{POD analysis and low-dimensional modeling}
\subsubsection{POD methodology}\label{POD}

The POD analysis of the unsteady flat plate wake is based on the commonly used  ``Method of Snapshots" proposed by \cite{siro87}, to find the dominant POD modes/eigenfunctions. The discrete data of the fluctuations of the velocity field are obtained from PIV measurements, and are arranged in vectors~\cite{Mand10, Pede02} as
\begin{equation}
\label{eqn:POD4}
\mathbf{\Psi}_i = \mathbf{V}_i - \frac{1}{M} \sum_{j=1}^{M} \mathbf{V}_j, \quad i=1,2,....,M ,
\end{equation}
where $M$ is the number of ensemble, and $\mathbf{V}_j$ is the instantaneous velocity field corresponding to the $j$th PIV realization. From these vectors of the velocity fluctuations, as the mean flow fields are removed, the elements of a covariance matrix are formed as,

\begin{equation}
\label{eqn:POD5}
R_{ij} = (\mathbf{\Psi}_i, \mathbf{\Psi}_j).
\end{equation}

The covariance matrix $R$ is a $M \times M$ symmetric matrix with non-negative eigenvalues, $\lambda$. These eigenvalues correspond to the energy of the respective POD mode. The energy fraction of the $k$-th mode is given by $E_k = \frac{\lambda_k}{E}$ where $E$ is the total energy of the POD modes. 

The eigenfunctions, $\mathbf{\Phi}^k $, are constructed using the eigenvectors of the co-variance matrix as,

\begin{equation}
\label{eqn:POD6}
\mathbf{\Phi}^k = \sum_{i=1}^{M} \mathbf{\phi}_i^k \mathbf{\Psi}_i,\quad k = 1,....,M,
\end{equation}
where $\phi_i^k$ is the $i$-th component component of the $k$-th eigenvector, and the eigenfunctions are normalized such that $(\mathbf{\Phi}^{k},\mathbf{\Phi}^{l})$ = $\delta_{kl}$, where $\delta_{kl}$ denotes the Kronecker delta~\cite{Mand10}. Using the eigenfunction system $\mathbf{\Phi}^{k}$, one can expand the fluctuating velocity field, $\mathbf{v}(\mathbf{x},t)$, as,
\begin{equation}
	\label{eqn:reco}
	\mathbf{v}(\mathbf{x},t_{n})=
	\sum_{k}a^{k}(t_{n})\mathbf{\Phi}^{k}(\mathbf{x}),
\end{equation}
where the time coefficients, $a^{k}(t_{n})$, are obtained by projecting the instantaneous snapshots on the eigenfunctions, i.e.
\begin{equation}
	\label{eqn:coeff}
	a^{k}(t_{n}) = (\mathbf{v}(\mathbf{x},t_{n}), \mathbf{\Phi}^{k}(\mathbf{x})).
\end{equation}

%It is worth mentioning that the POD modes occur in pairs, which is attributed to their ability to capture the coherent structures present in the flows \cite{remp94}.

\subsubsection{Low-dimensional modeling}\label{LOD}
Associated with the Galerkin projection, low-dimensional modeling method allows one to project the Navier--Stokes equations onto the orthonormal POD modes and get a system of ordinary differential equations for the time coefficients, $a_{k}(t)$. For Galerkin projection, we follow the formulation presented in~\cite{raja94, indr18}. Since POD eigenfunctions satisfy the continuity equation, we consider the following momentum equation for the instantaneous velocity, $\mathbf{V}$. Neglecting the body force term and considering incompressible flow, the momentum equation for the instantaneous velocity, $\mathbf{V}$, reads as
\begin{equation}
	\label{eqn:momen}
	\frac{\partial \mathbf{V}}{\partial t} + \mathbf{V}\cdot\nabla\mathbf{V}  = -\frac{\nabla P}{\rho} + \nu \Delta \mathbf{V},
\end{equation}

\noindent where $P$ is the pressure and $\nu$ is the kinematic viscosity; here, $\mathbf{\overline{V}}$  and $\mathbf{v}$ are the mean and the fluctuating parts of the velocity. Substituting $\mathbf{V}=\overline{\mathbf{V}}+\mathbf{v}$  in equation~(\ref{eqn:momen}) and subtracting the time-averaged momentum equation, we obtain
\begin{equation}
	\label{eqn:fleq}
	\frac {\partial \mathbf{v}}{\partial t}+ \mathbf{v}\cdot 
	\mathbf{\nabla \overline{V}}+\mathbf{\overline{V}}\cdot  \mathbf{\nabla
		v}+\mathbf{v}\cdot  \mathbf{\nabla v}-\overline{\mathbf{v}\cdot  \mathbf{\nabla v}}= -
	\frac{1}{\rho}\mathbf{\nabla} p + \nu \mathbf{\triangle}\mathbf{v},
\end{equation}
where $p$ is the fluctuating pressure. Using the Eq.~\ref{eqn:reco} in the above, and projecting the resulting equation on the POD modes, $\Phi_k$, we have
\begin{equation}
	\label{eqn:dyn}
	\frac{da^{k}}{dt}=A^{ki}a^{i}+B^{kil}(a^{i}a^{l}-\overline{a^{i}a^{l}})+C^{k},
\end{equation}
\noindent where the coefficients are given by 
\begin{equation}
	A^{ki}=-(\mathbf{\Phi}^{k},\mathbf{\Phi}^{i}\cdot \mathbf{\nabla \overline{V}})-(\mathbf{\Phi}^{k},\mathbf{\overline{V}}\cdot \nabla \mathbf\Phi^{i})+ \nu (\mathbf\Phi^k,\triangle \mathbf\Phi^{i}),
\end{equation}
\begin{equation}
	B^{kil}=-(\mathbf\Phi^k,\mathbf{\Phi}^{i}\cdot \nabla \mathbf{\Phi}^{l}), 
\end{equation}
\begin{equation}
	 C^{k}= -\frac{1}{\rho}(\mathbf\Phi_k, \nabla p).
\end{equation}
Note that $\overline{a^{k}a^{l}}$ = $\delta_{kl}\lambda_{k}$ as the coefficients are uncorrelated. Utilizing the given POD modes and the mean velocities, one can calculate the coefficients of Eq.~\ref{eqn:dyn}. Further, one can find that $C^{k} = 0$, considering homogenous boundary conditions and divergence free nature of the POD modes (see~\cite{raja94} for further details).

\section{Mean flow fields and shedding frequency}

 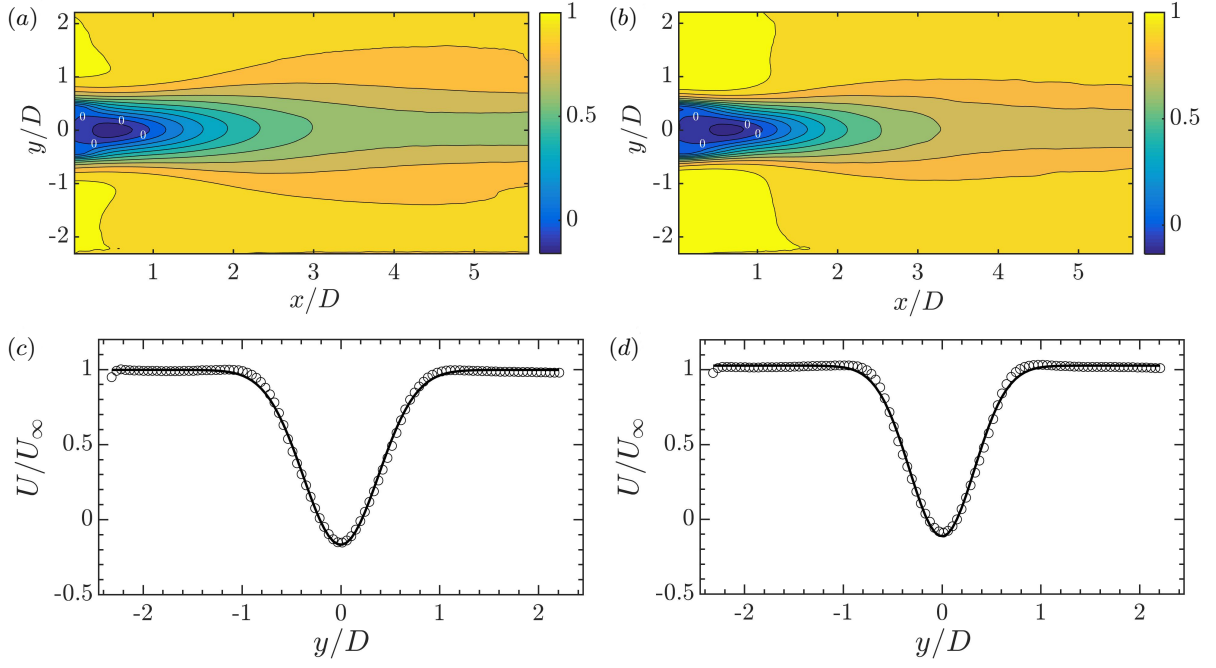
\begin{figure}
 \unitlength=40.0mm
 \unitlength=40.0mm
\centerline{
\includegraphics[width=1.967\unitlength,height=1.112\unitlength]{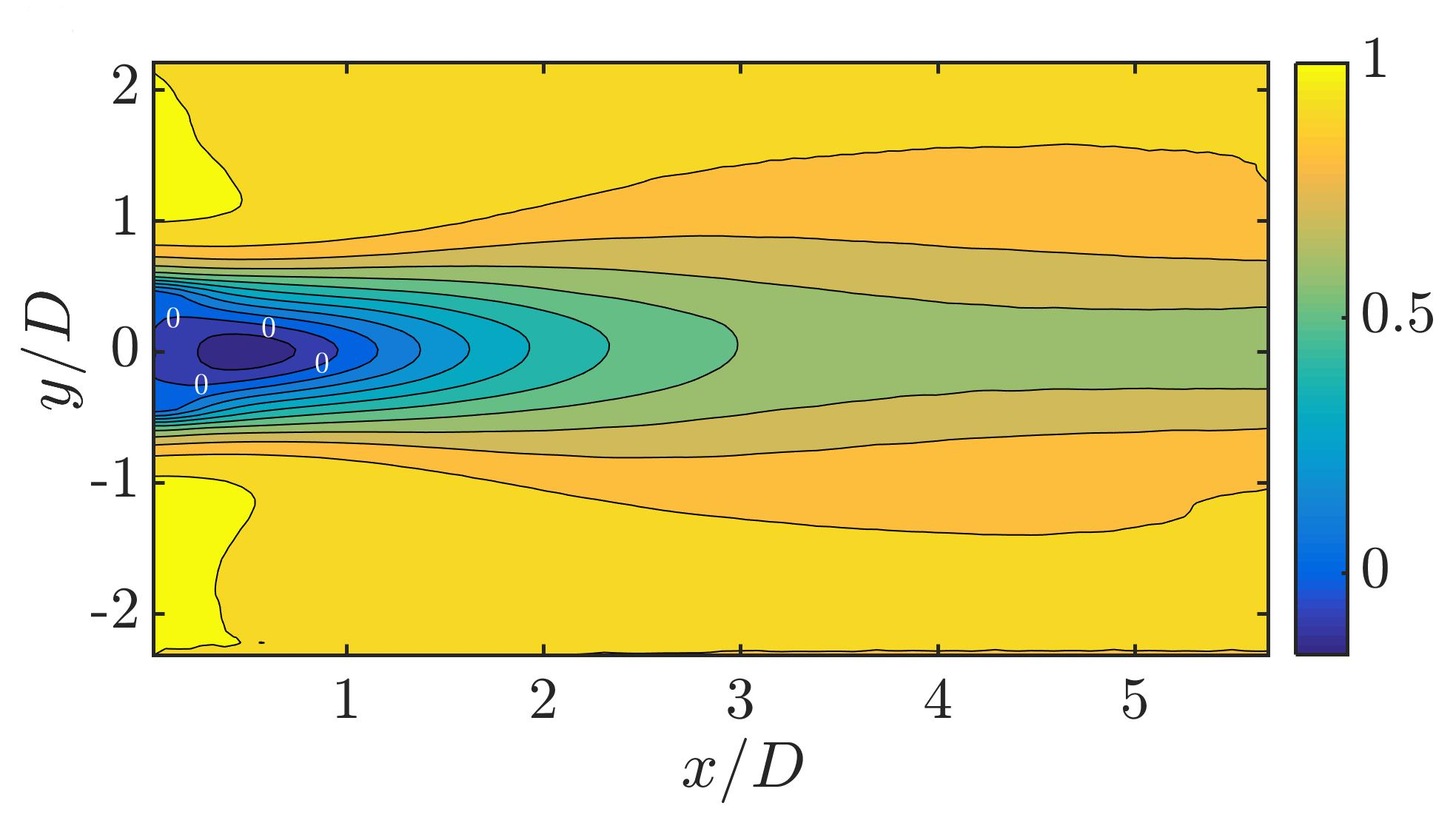}
\includegraphics[width=1.967\unitlength,height=1.112\unitlength]{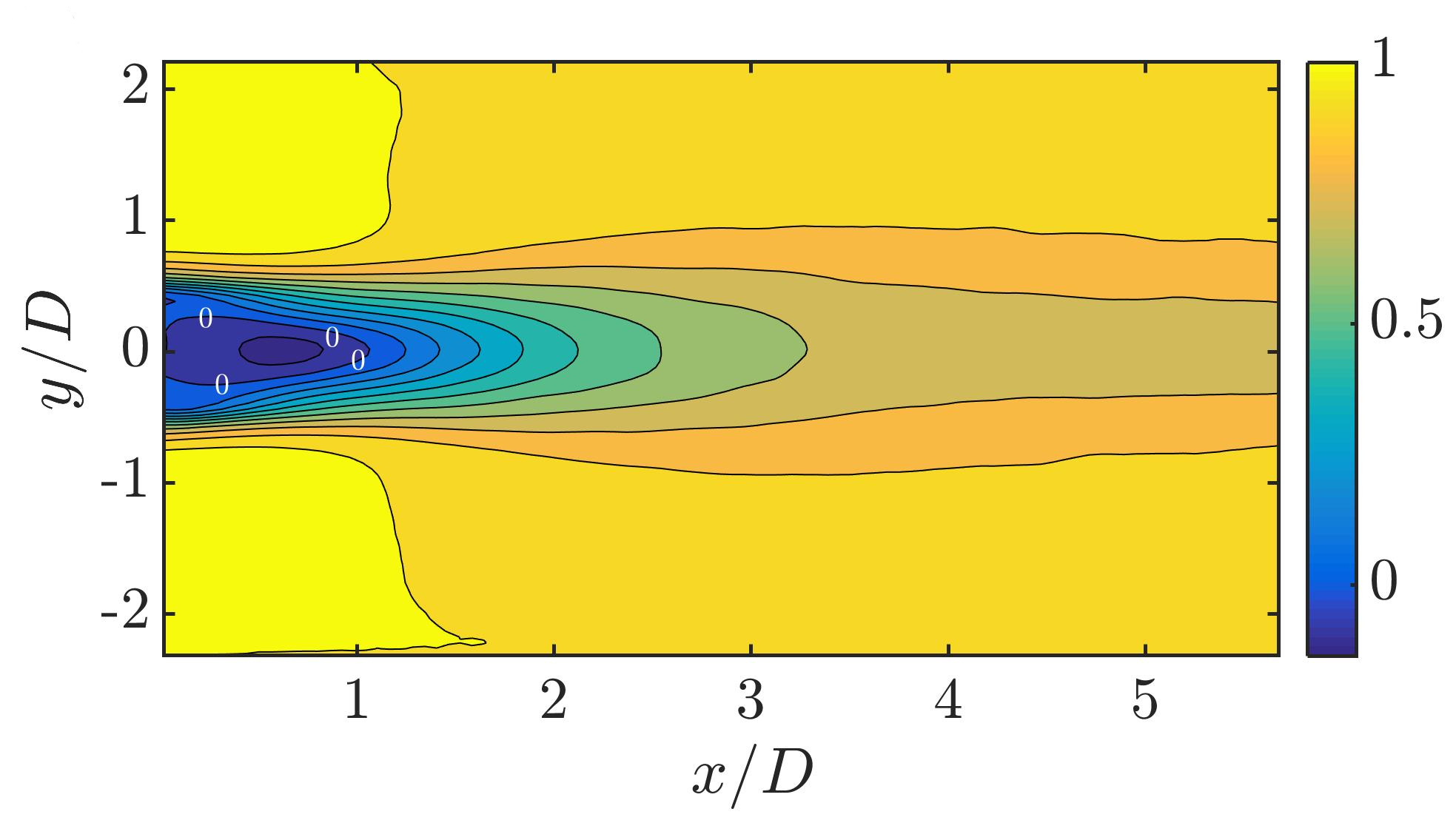}
}
\centerline{
\includegraphics[width=1.967\unitlength,height=1.112\unitlength]{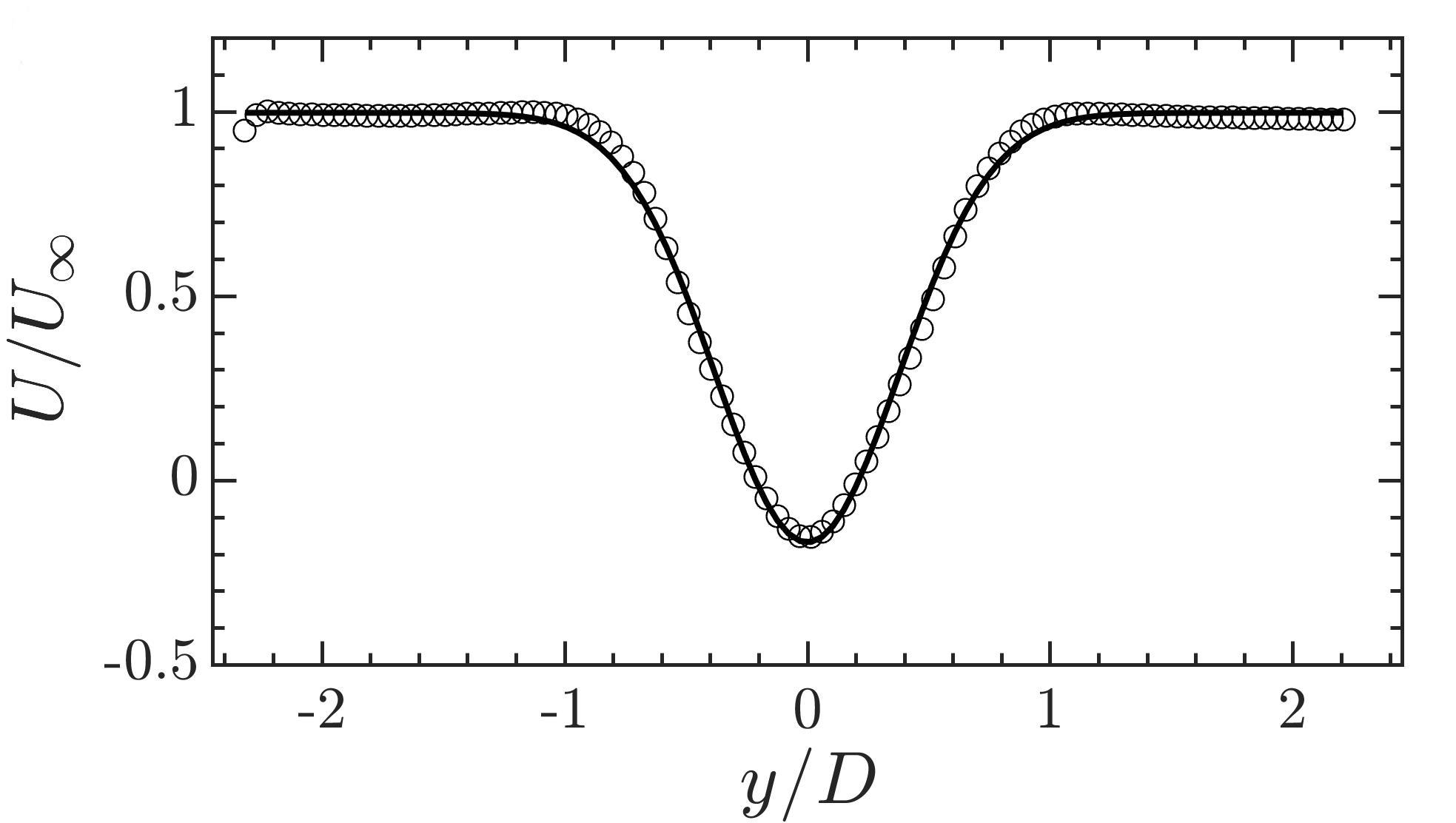}
\includegraphics[width=1.967\unitlength,height=1.112\unitlength]{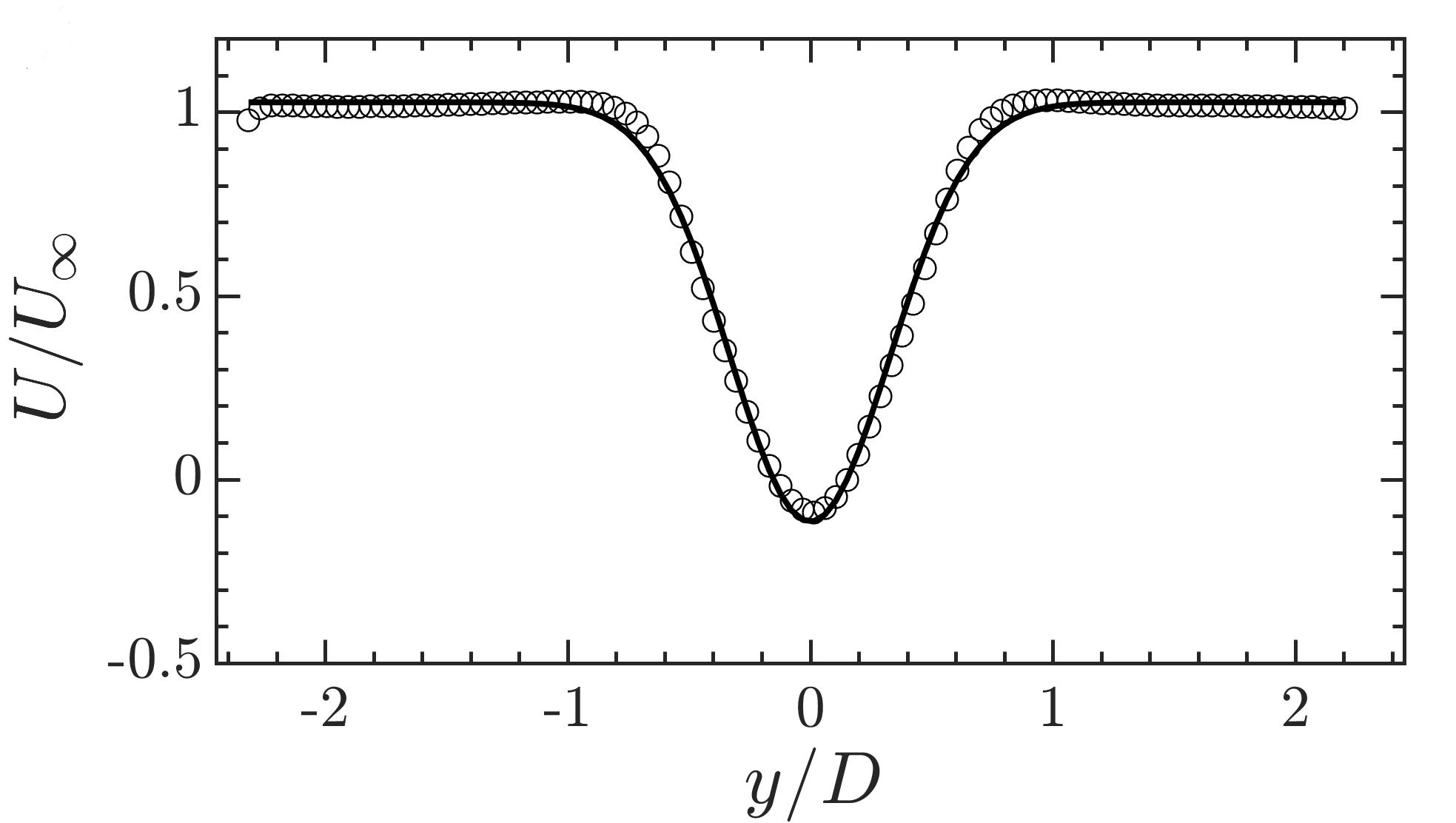}
}
\begin{picture}(0,0)
\put(-2.0,2.2){${(a)}$}
\put(0.0,2.2){$(b)$}
\put(-2.0,1.1){$(c)$}
\put(0.0,1.1){$(d)$}
\end{picture}
 \caption{Mean flow fields of the wake of a flat plate with a blunt trailing edge. 
The spatial distribution of the time-averaged mean flow fields of streamwise velocity component for $(a)$ $Re = 1850$ and $(b)$ $Re = 3350$. The zero contour represents the extent of the recirculation zone. The wall-normal mean velocity profile $U(y) = U_b(x,y)$ at $(a)$ $x/D = 0.86$ for $Re = 1850$, and $(b)$ $x/D = 0.56$ for $Re=3350$. The continuous lines represent the best fit mean flow velocity profiles given by Eq.(\ref{eqnvelfit}) which are used in the local spatio-temporal stability analyses, with $(c)$ $c_1 = -1.165, c_2 = 0.289, c_3 = 0.996$, and $(d)$ $c_1 = -1.141, c_2 = 0.217, c_3 = 1.026$.
}
 \label{fig:velprofiles}
 \end{figure} 

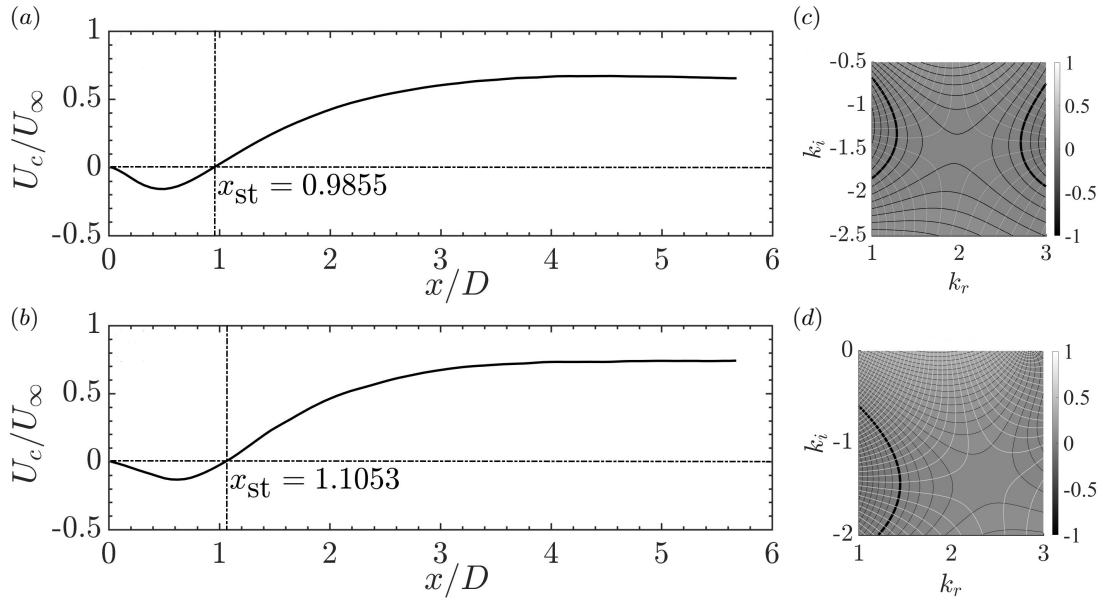
\begin{figure}
 \unitlength=40.0mm
\centerline{
\includegraphics[width=2.616\unitlength,height=0.97\unitlength]{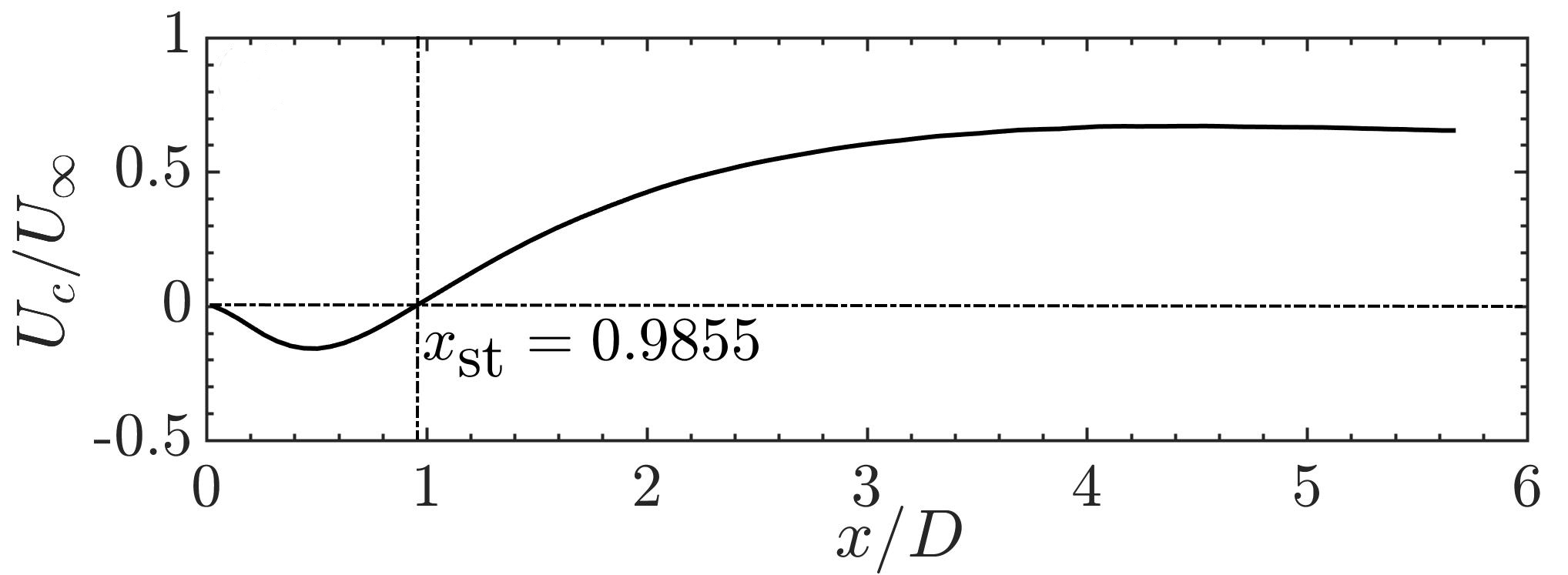}
\includegraphics[width=0.985\unitlength,height=0.875\unitlength]{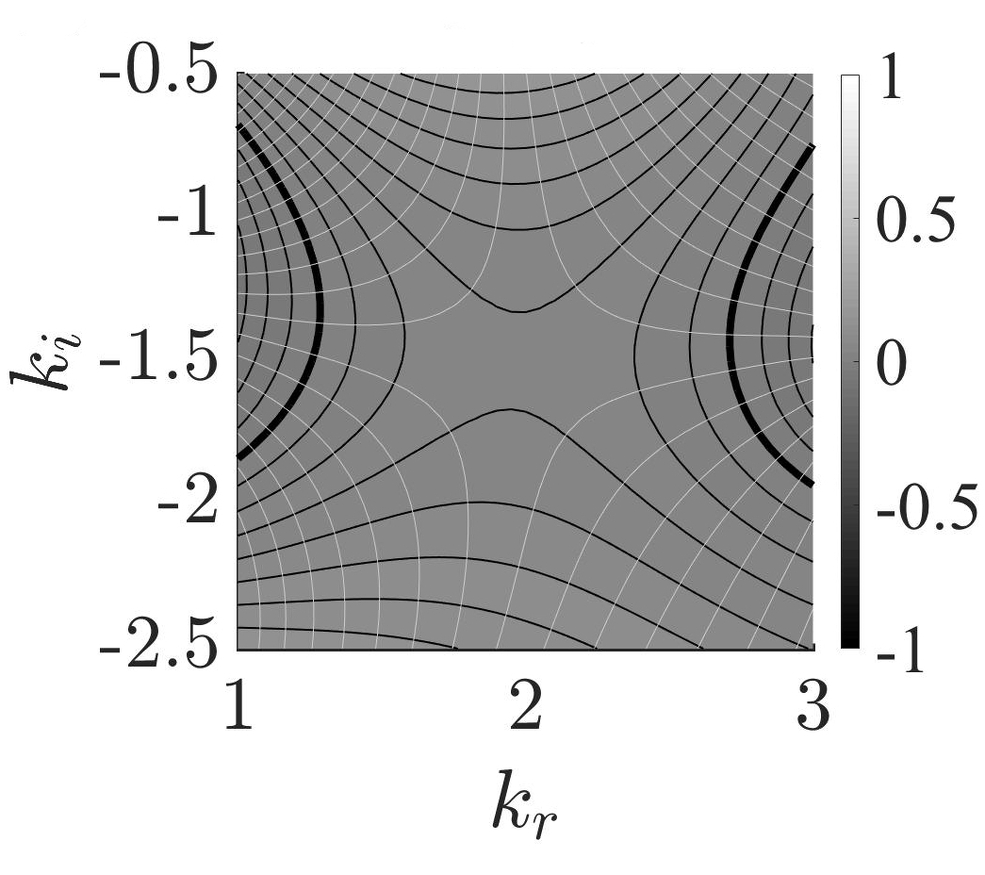}
}
\centerline{
\includegraphics[width=2.616\unitlength,height=0.97\unitlength]{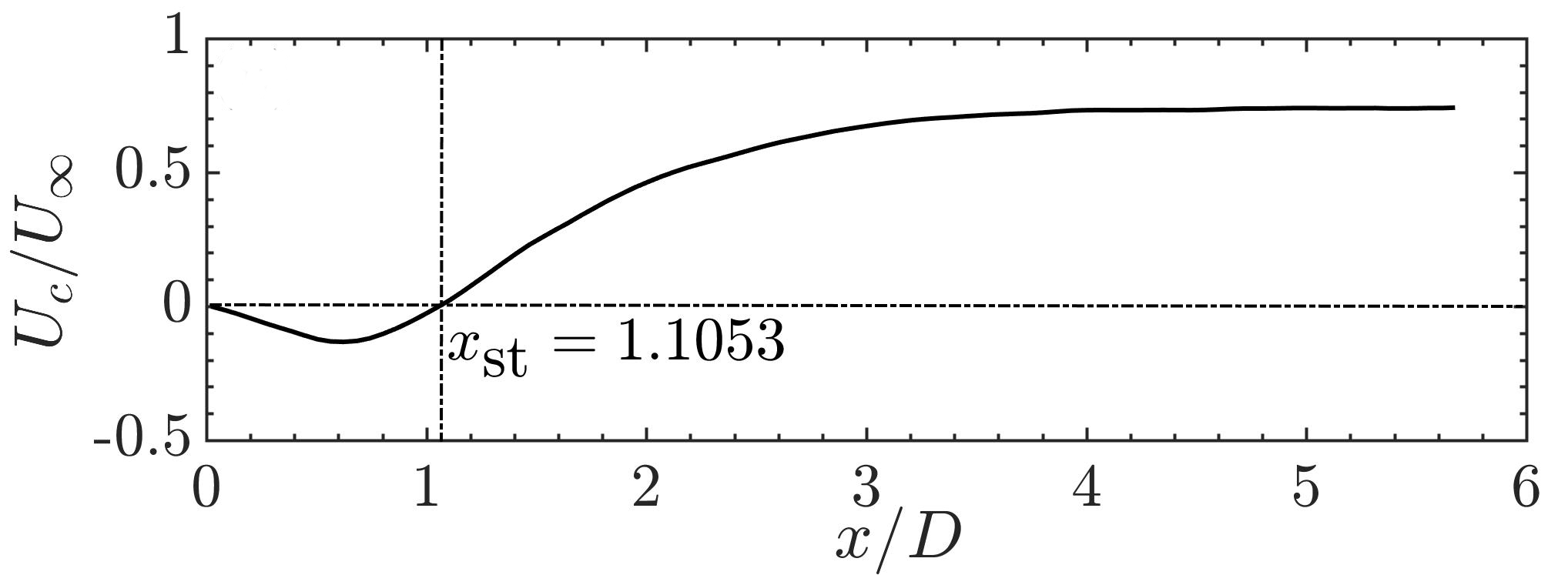}
\includegraphics[width=0.985\unitlength,height=0.875\unitlength]{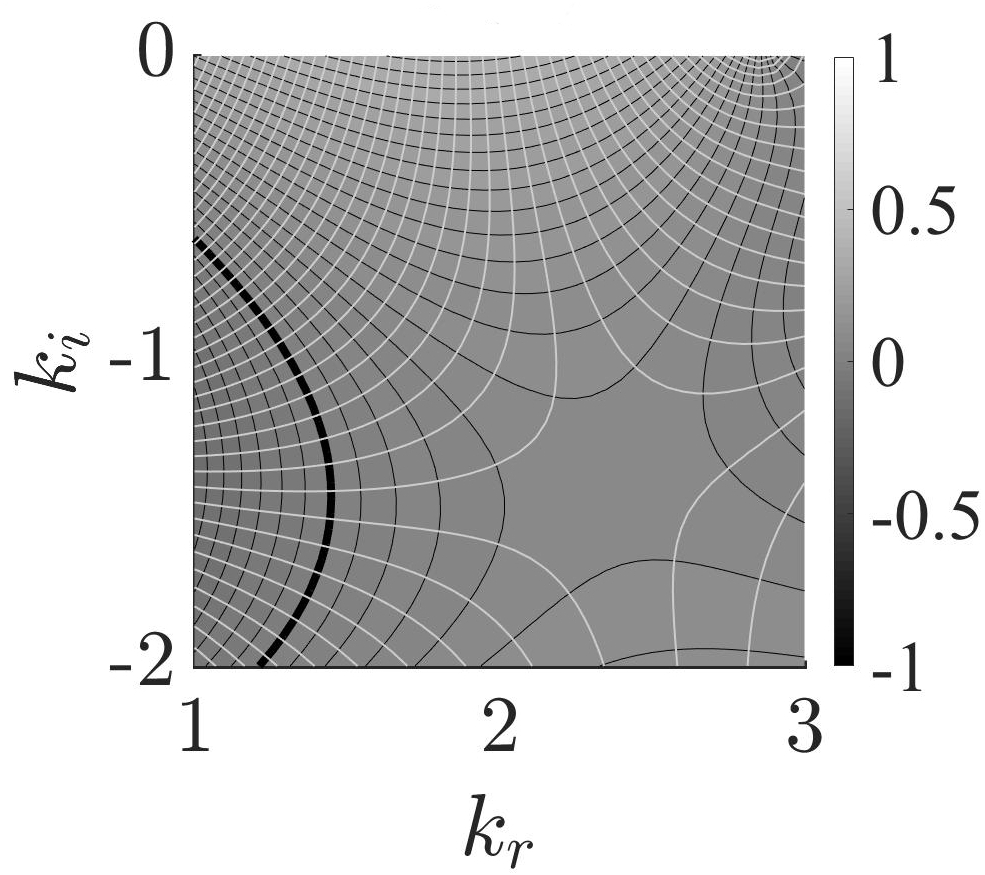}
}
\begin{picture}(0,0)
\put(-1.8,2.0){$(a)$}
\put(0.8,2.0){$(c)$}
\put(-1.8,1.0){$(b)$}
\put(0.8,1.0){$(d)$}
\end{picture}
 \caption{Variation of the centerline velocity as a function of the downstream distance behind the blunt trailing edge of the flat plate.  Parameter settings: (a) $Re = 1850$, (b) $Re = 3350$. Contours of $\omega_{\dot{\imath}}$ (greyscale) and $\omega_r$ (greylines) at $Re = 1850$ and $Re = 3350$ at the corresponding stagnation points. The thick black line denotes $\omega_i = 0$.
}
 \label{fig:spattemp1}
 \end{figure}

We now present the mean velocity fields of the flat plate wake obtained for blunt, and circular, trailing edges for two values of the Reynolds number. Fig.\ref{fig:velprofiles} shows the time-averaged velocity profiles of the flat plate wake with a blunt trailing edge. The spatial distribution of the mean flow fields, $U_b(x,y)$, are presented in Fig. \ref{fig:velprofiles}$(a,b)$. Directly downstream of the trailing edge, the flow reverses, as can be seen from the negative values 
of the centerline velocity. Moving downstream along the flat plate wake, we can observe that the centerline velocity 
decreases up to a point where the flow reversal is maximum, following which it increases. 
The variation of the mean flow velocity along the wall-normal direction at two different streamwise locations are shown in 
panels $(c,d)$ for $Re = 1850$ and $3350$, respectively. To perform a local stability analysis of these time-averaged velocity fields, as in other experimental \cite{asai2002instability,Bala17} and numerical studies \cite{julien2004secondary}, it is convenient to fit the measured profiles using an analytical expression as
\begin{equation}
U(y) = c_1 \exp(-y^2/c_2) + c_3.
\label{eqnvelfit}
\end{equation}
Here $c_1$, $c_2$, and $c_3$ are constants which were varied so as to best predict the measured experimental velocity profiles. Indeed, their values are different at each streamwise location. These are shown by the solid black lines in Fig.\ref{fig:velprofiles}$(c,d)$. The length of the recirculation zone, represented using the zero contour of the streamwise velocity field in panels $(a,b)$, increases with the Reynolds number. 
This is elucidated in Fig.\ref{fig:spattemp1} where the variation of the centerline velocity downstream of the flat plate with a blunt trailing edge are shown in panels $(a,b)$. The contours of the complex frequency $\omega$ in the complex $k$ plane at the respective stagnation points for $Re = 1850$ and $Re = 3350$ are shown in panels $(c,d)$.

 \begin{figure}
 \unitlength=36.0mm
\centerline{
\includegraphics[width=1.967\unitlength,height=1.112\unitlength]{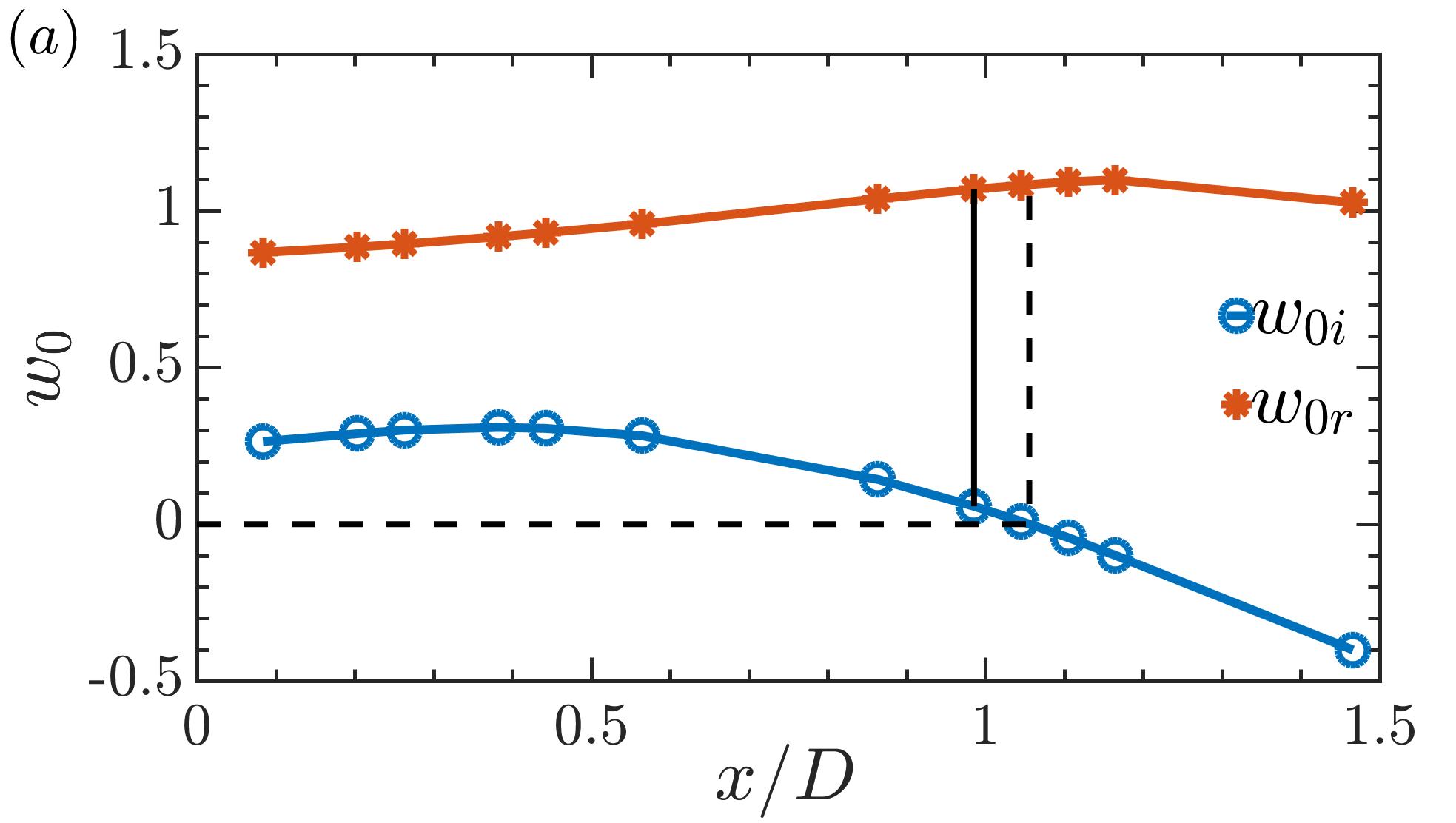}
\includegraphics[width=1.967\unitlength,height=1.112\unitlength]{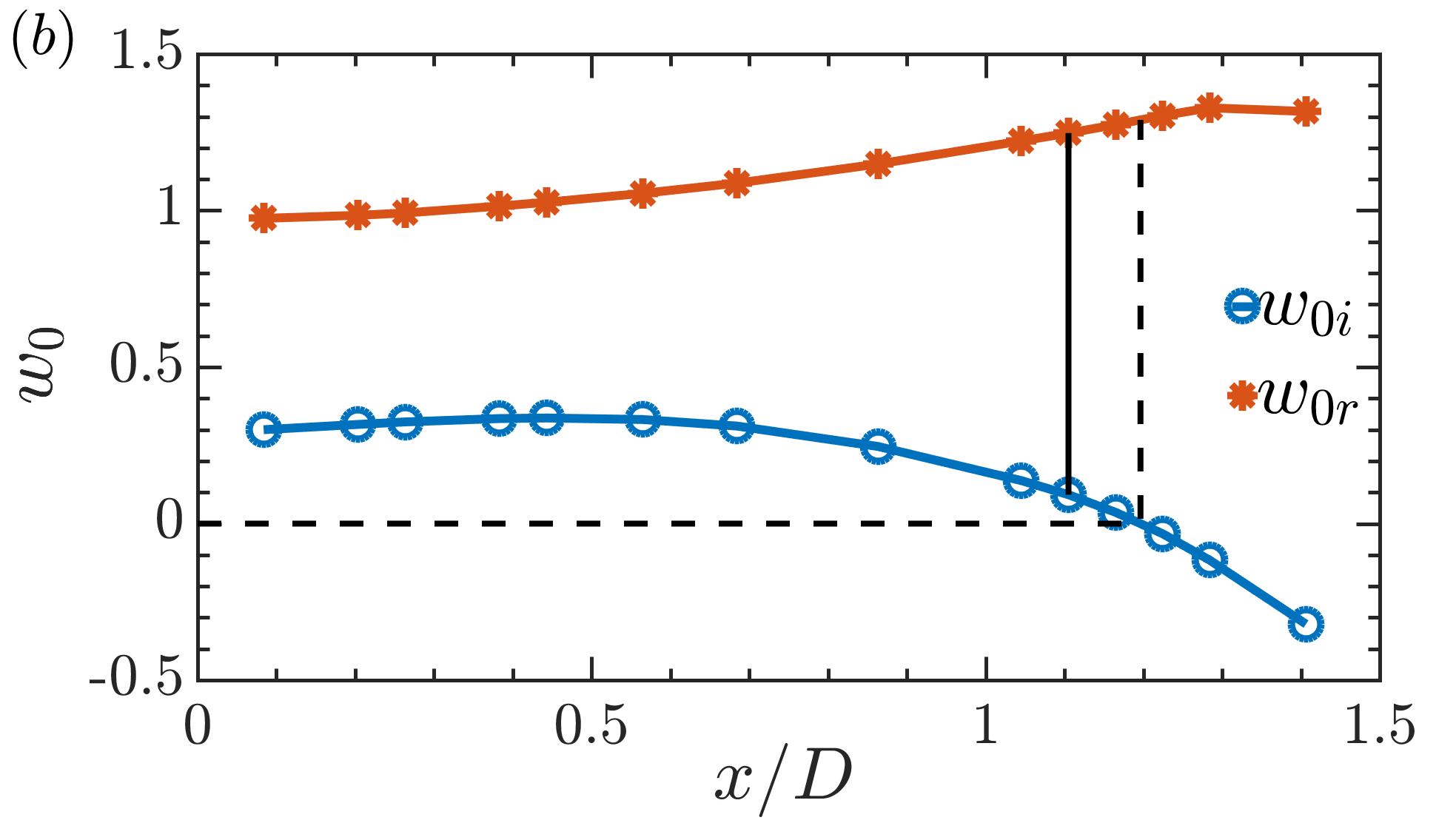}
}
\centerline{
\includegraphics[width=1.967\unitlength,height=1.112\unitlength]{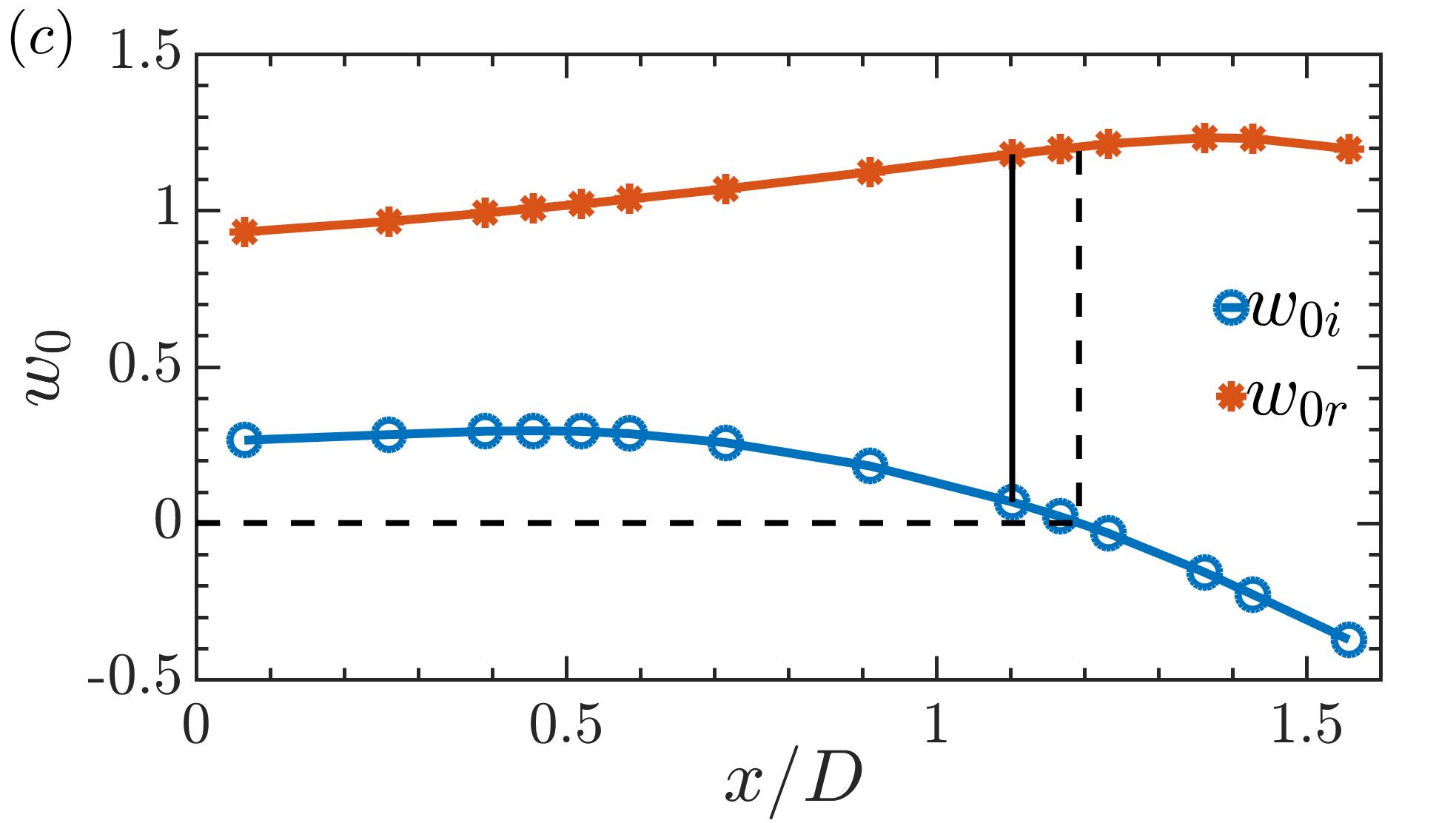}
\includegraphics[width=1.967\unitlength,height=1.112\unitlength]{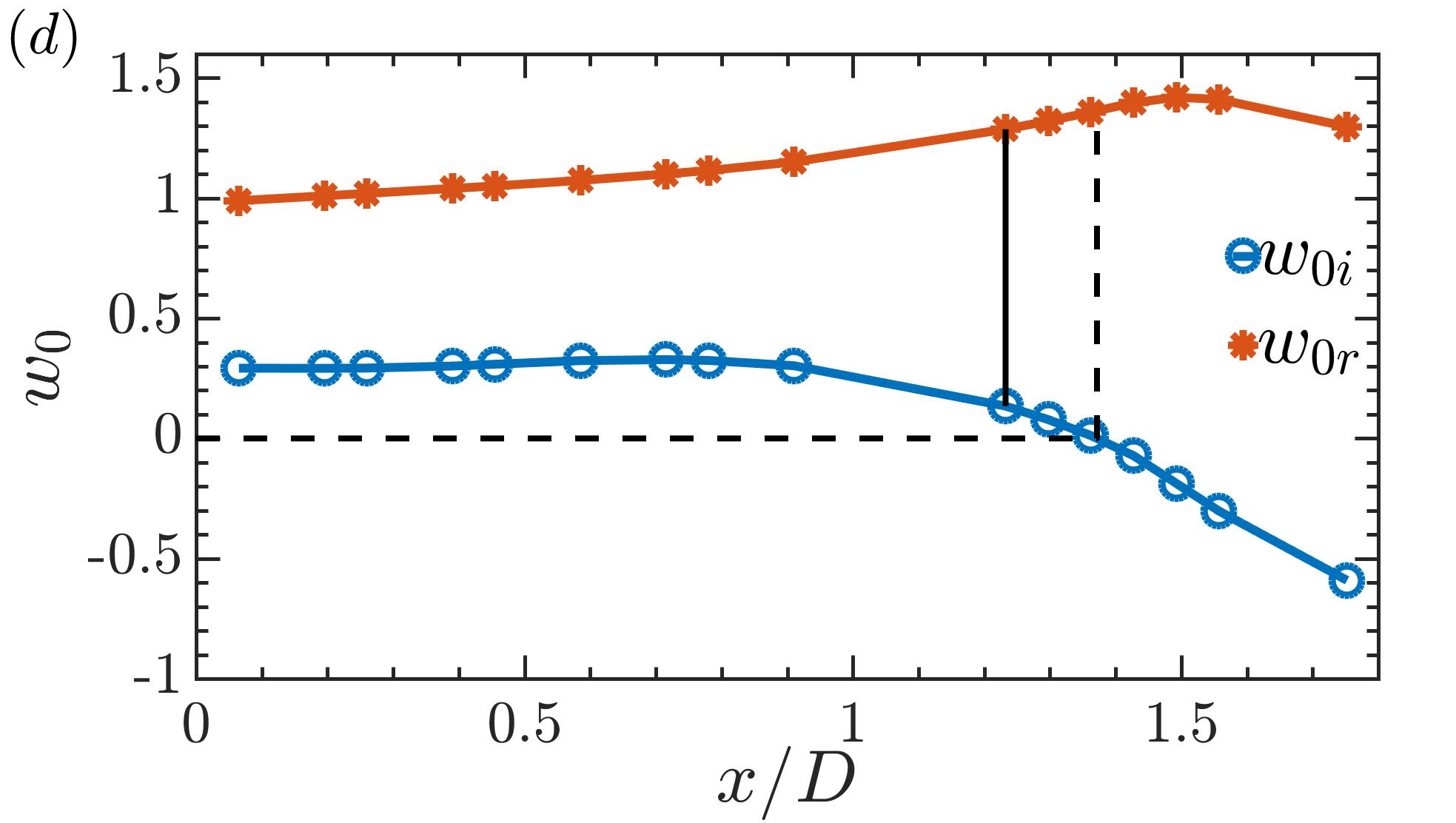}
}
 \caption{Temporal amplification rate $\omega_{0i}$ (circles) and $\omega_{0r}$ (diamonds) as a function of the streamwise distance 
for the different flat plate wakes in the present study. The upper panels $(a,b)$ correspond to the blunt trailing edge, while $(c,d)$ for the circular trailing edge. The straight vertical line represents the stagnation point $x_{\textrm{st}}$, while the dashed line indicates the end of the AU region. Parameter settings: $(a,c)$ $Re = 1850$ and $(b,d)$ $Re = 3350$.
}
 \label{fig:absomzero}
 \end{figure}

To delineate the absolutely unstable (AU) region from the convectively unstable (CU), we extend the local analysis to the entire wake region. Fig.\ref{fig:absomzero} 
shows the variation of the real and imaginary parts of the absolute frequency $\omega_0$ as a function of the downstream distance from the trailing edge of the flat plate. The AU region where $\omega_{0i} > 0$ is demarcated using dotted lines. 
It can be seen in Fig.\ref{fig:absomzero} that the flow is locally AU both for blunt, and circular trailing edges directly downstream in the flat plate wake. The extent of the AU region indeed increases with the Reynolds number, beyond which the flows are CU. 
It is this boundary of the AU region which is of particular interest, as they are defined by real absolute frequencies \cite{pier1998steep,Pier1}.
It can be seen from Fig.\ref{fig:absomzero} that the absolutely unstable region closely follows the flow reversal region, extending slightly beyond it. For $Re = 1850$ the distance between the stagnation point and the front of the AU region is $0.0705D$, while for $Re = 3350$ is $0.0912D$. 
This feature has also been observed in other bluff-body wake flows \cite{Pier1, Pier2, Hwang1, Hammond1}.
It can also be noted from Fig.\ref{fig:absomzero} that the local frequency at a given streamwise location in the wake is higher at a larger Reynolds number.

   \begin{figure}
 \unitlength=70.0mm
\centerline{
\includegraphics[width=1.967\unitlength,height=1.112\unitlength]{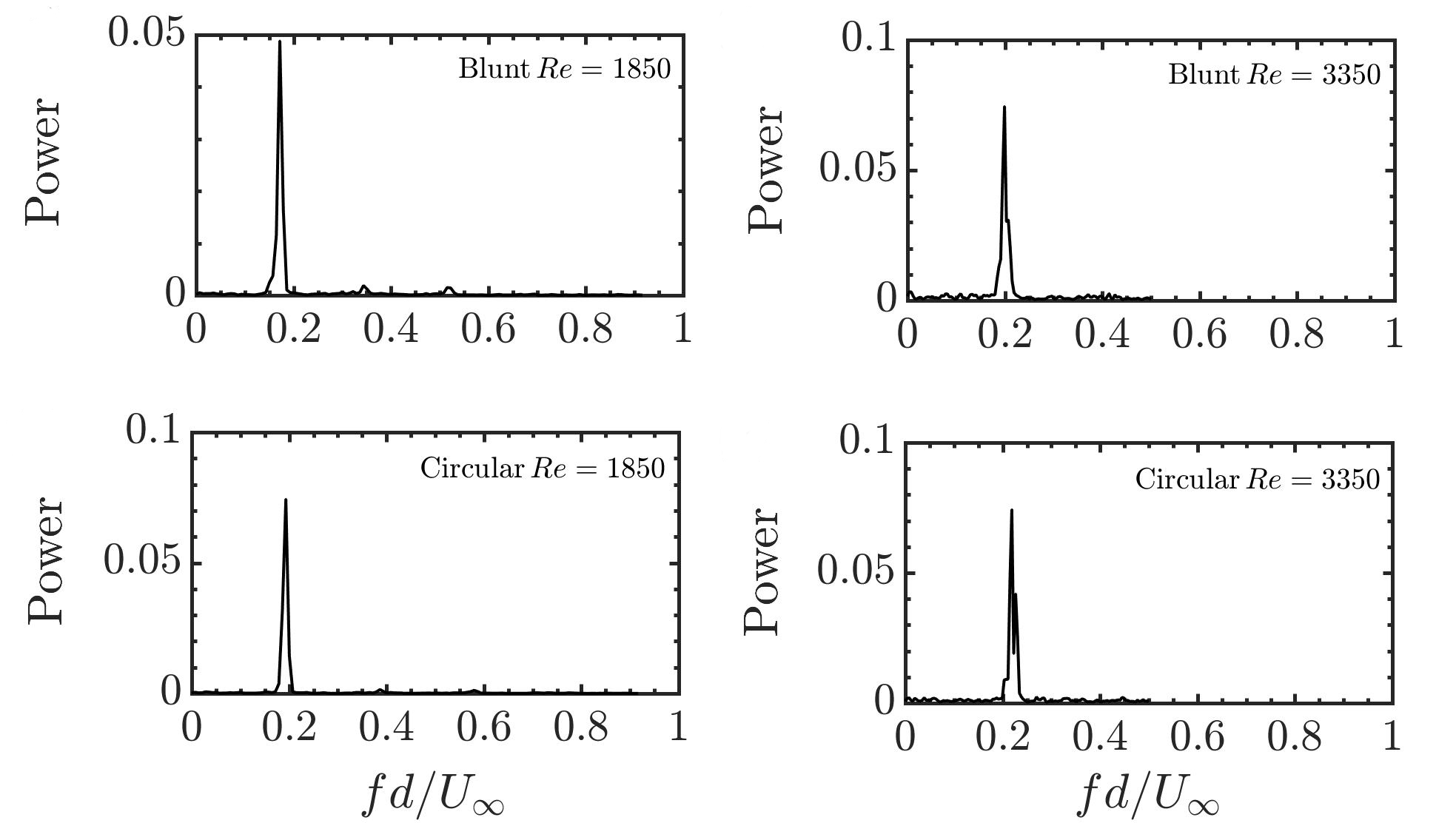}
}
\begin{picture}(0,0)
\put(-1.0,1.15){$(a)$}
\put(0.0,1.15){$(b)$}
\put(-1.0,0.6){$(c)$}
\put(0.0,0.6){$(d)$}
\end{picture}
 \caption{The experimentally measured frequency spectra for the different cases considered in the present study. The global shedding frequency increases with the Reynolds number, and remains roughly the same irrespective of the trailing edge geometry.
}
 \label{fig:freqexp}
 \end{figure}
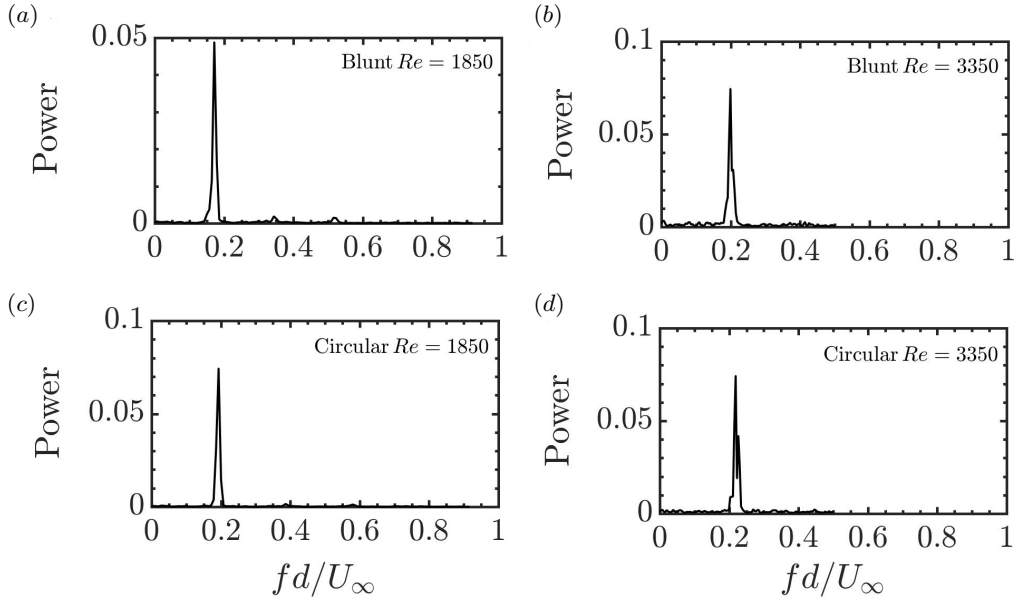 

% \begin{table}
% \caption{Summary of the different characteristic frequencies (in Hz) from the present study. Here $f_{\textrm{st}}$ and $f_{\textrm{AU/CU}}$ are the frequencies obtained from a local spatio-temporal analysis at the stagnation point and at the end of AU region respectively. The experimentally measured shedding frequency is denoted by $f_{\textrm{exp}}$, and $f_{\textrm{POD}}$ corresponds to the POD analysis.}
% \begin{tabular}{|c|c|c|c|c|c|c|}
% \tableline 
% Trailing edge & $Re$ &  $f_{\textrm{st}}$ & $f_{\textrm{AU/CU}}$ & $f_{\textrm{exp}}$ &$f_{\textrm{saddle}}$ & $f_{\textrm{POD}}$\\
% \tableline
%Blunt & $1850$ &  $33.9$ & $34.35$ & $34.22$& 29.9 & 33.48\\
% \tableline
%Blunt & $3350$ &  $71.63$ & $74.04$ & $72.71$& 60 & 72.5\\
% \tableline
%Circular & $1850$ &  $37.40$ & $38.15$ & $38.50$& 32.34 &39.2 \\
% \tableline
% Circular & $3350$ & $73.90$  & $78.31$ & $79.84$& 63.5 & 78.24 \\
% \tableline
% \end{tabular}
% \label{tab:FreqChar}
% \end{table}

\begin{table}
		\begin{tabular}{|c|c|c|c|c|c|c|}
		\tableline 
		Trailing edge & $Re$ &  $\omega_{\textrm{st}}$ & $\omega_{\textrm{AU/CU}}$ & $\omega_{\textrm{exp}}$ &$\omega_{\textrm{saddle}}$ & $\omega_{\textrm{POD}}$\\
		\tableline
		Blunt & $1850$ &  $1.0425$ & $1.0549$ & $1.0487$& $0.9184$ & $1.0301$\\
		\tableline
		Blunt & $3350$ &  $1.2331$ & $1.2745$ & $1.2516$& $1.0329$ & $1.2480$\\
		\tableline
		Circular & $1850$ &  $1.1653$ & $1.1886$ & $1.1995$& $1.0076$ & $1.2214$ \\
		\tableline
		Circular & $3350$ & $1.2721$  & $1.3480$ & $1.3744$& $1.0931$ & $1.3468$ \\
		\tableline
	\end{tabular}
	\caption{Summary of the different characteristic frequencies ($\omega_{0r}$) from the present study. Here $\omega_{\textrm{st}}$ and $\omega_{\textrm{AU/CU}}$ are the frequencies obtained from a local spatio-temporal analysis at the stagnation point and at the end of AU region respectively. The experimentally measured shedding frequency is denoted by $\omega_{\textrm{exp}}$. Here, $\omega_{\textrm{saddle}}$ denotes the circular frequency obtained from the saddle point criterion, and $\omega_{\textrm{POD}}$ obtained from the POD based low dimensional model, as described below. }
	\label{tab:FreqChar}
\end{table}

It has to be pointed out that unlike in cylinder wakes, the flow is AU at the trailing edge of the flat plate. In the former the flow goes from a locally CU domain to a finite sized region where the flow is locally AU. The frequency at this transition station, from local convective to absolute instability,
fixes the global shedding frequency \cite{Pier2}. This arises from a balance between 
upstream perturbation growth from the AU region and downstream advection from the CU region resulting in a stationary front \cite{dee1983propagating,Pier2}. However, in flat plate wakes the flow is already AU at the trailing edge, and hence the global selection frequency outlined in \cite{Pier2} is not directly applicable to the present flows. We shall see that the other characteristic frequencies discussed in \cite{Pier2}, notably the real absolute frequency corresponding to the AU/CU transition station \cite{Pier1}, which corresponds to the criterion outlined in \cite{koch1985local}, and the local absolute frequency at the stagnation point, predicts the observed global shedding frequency. The dimensional absolute frequency at any location can be calculated as
%\large{
%\begin{equation}
%f = \frac{\omega_{0r}}{2\pi}\Big(\frac{U_{\infty}}{D}\Big).
%\end{equation}
%}
\large{
\begin{equation}
\omega_{0r} = 2\pi \Big(\frac{fD}{U_{\infty}}\Big).
\end{equation}
}
\normalsize
For $Re = 1850$ the free stream velocity $U_{\infty}$ is $2.42\textrm{m/s}$, whereas for $Re = 3350$, $U_{\infty}$ is $4.38\textrm{m/s}$. The flat plate thickness $D$ in the current study is $12\textrm{mm}$. The characteristic frequencies at some streamwise locations of interest, along with the experimentally measured values are presented in table \ref{tab:FreqChar}. 
Fig. \ref{fig:freqexp} summarises these experimentally measured global shedding frequencies in the present study. They are in excellent agreement with the local absolute frequency at the streamwise location where the flow goes from an absolutely unstable to a convectively unstable region. It may be noted here that the circular frequency, $\omega_{\textrm{saddle}}$, is obtained based on the analytical continuation of $\omega_{0}$ in the complex $x$ plane \cite{chomaz2005global,Hammond1} following a second order polynomial fit through $\omega_{0}$ as described in \cite{Rees2009}. To further investigate this excellent agreement using the time-averaged velocity profiles, we perform the POD analysis of the fluctuating velocity data; $\omega_{\textrm{POD}}$ in table \ref{tab:FreqChar} is found from this analysis, and is presented in the following section.

\section{Low dimensional modeling}\label{modeling}

 \begin{figure}
 \unitlength=50.0mm
\centerline{
\includegraphics[width=1.967\unitlength,height=1.112\unitlength]{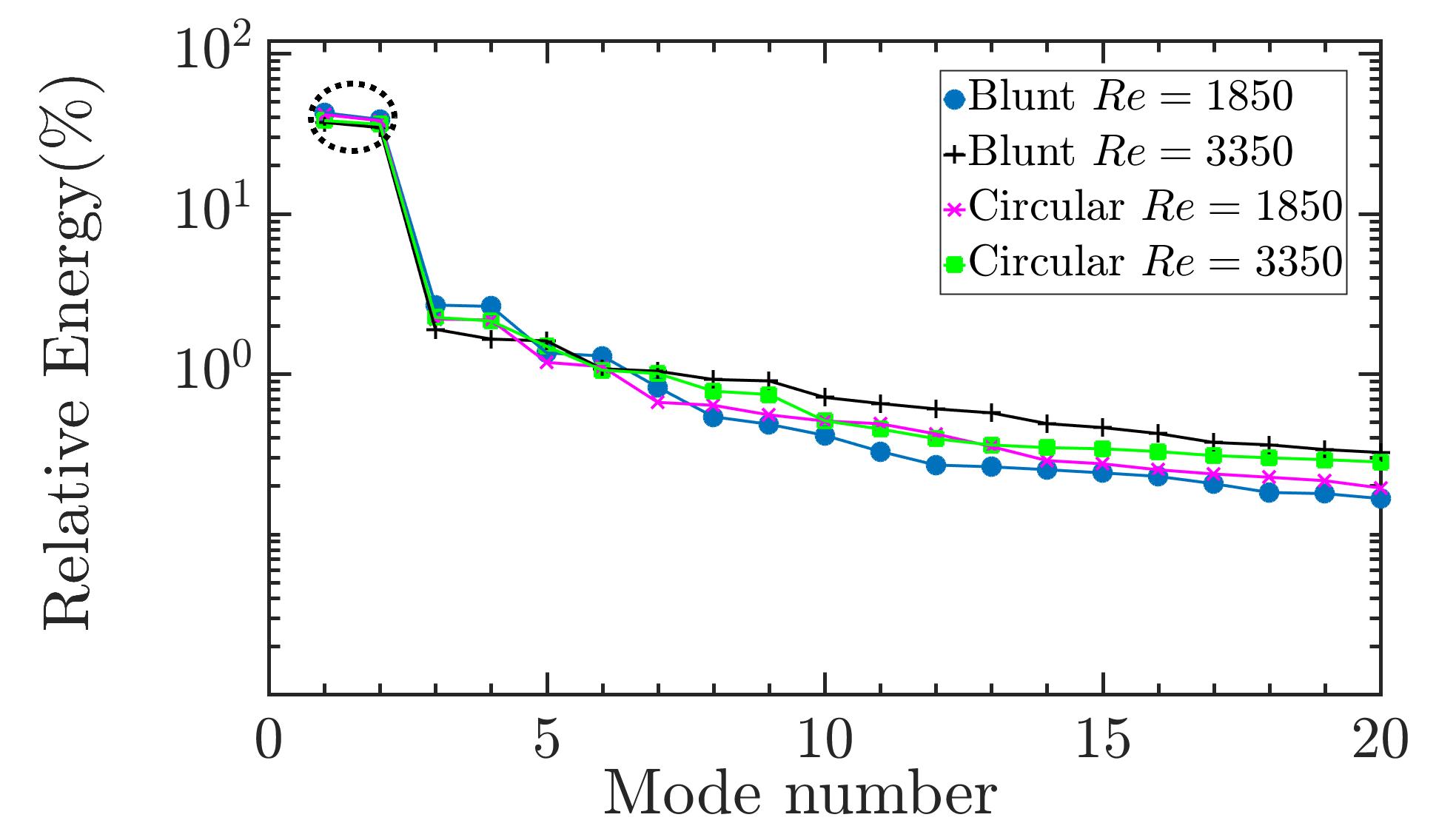}
}
 \caption{The relative contribution of the POD modes to the total energy. It can be seen that the first two POD modes, which occur in pairs, contribute over $80\%$ to the total energy in all the four different cases considered in the present study. 
}
 \label{fig:pod1}
 \end{figure}

  \begin{figure}
 \unitlength=95.0mm
\centerline{
\includegraphics[width=1.72\unitlength,height=1.1\unitlength]{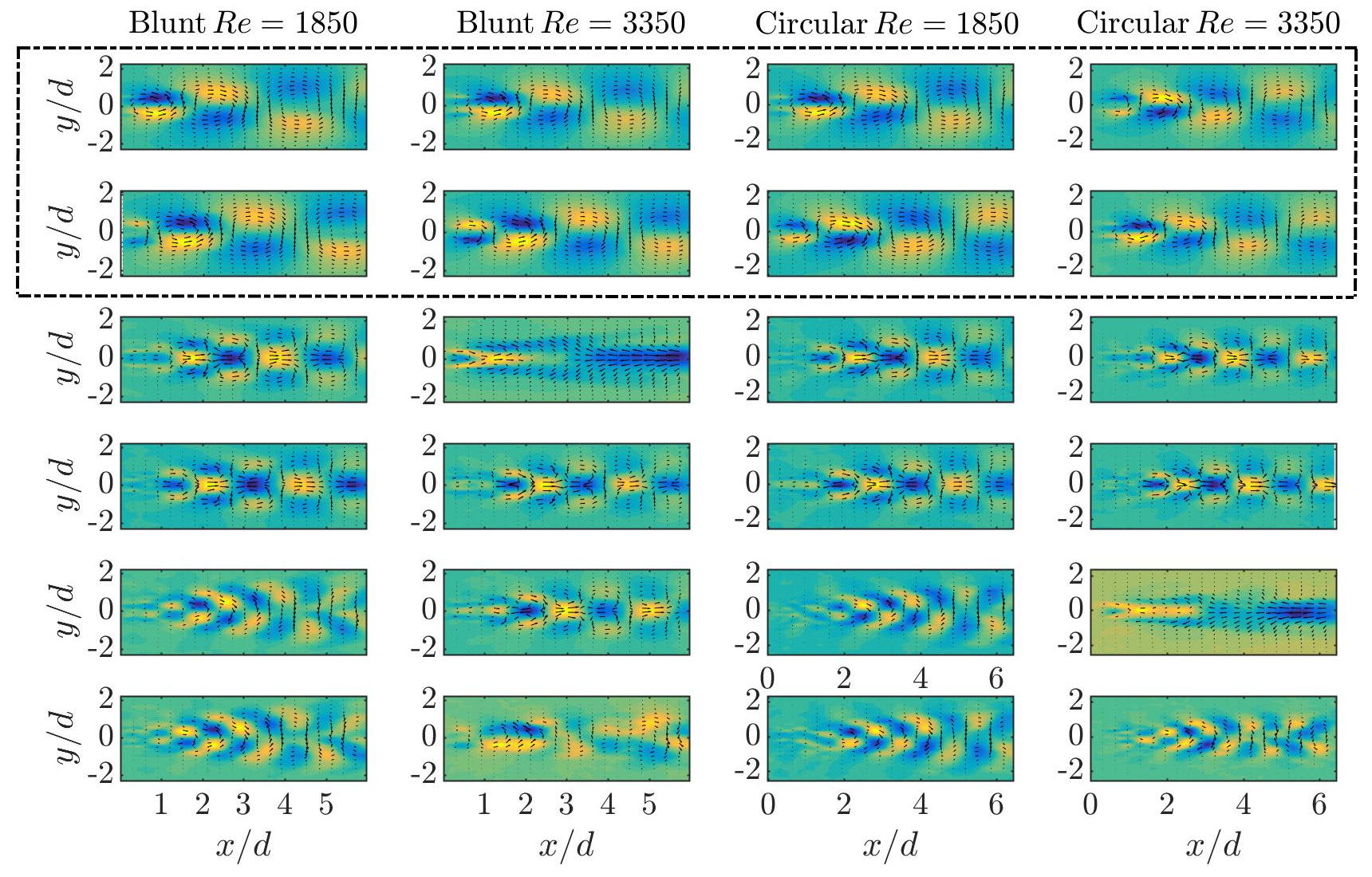}
}
 \caption{The spatial distribution of the first six POD modes for the four different cases considered in the present study. It can be seen that the first two modes have the same spatial structure, albeit with a slight shift in the streamwise direction. The fact that the POD modes occur in pairs is also manifestly evident from their spatial distributions.
}
 \label{fig:pod2}
 \end{figure} 

  \begin{figure}
	\unitlength=95.0mm
	\centerline{
		\includegraphics[width=1.85\unitlength,height=1.15\unitlength]{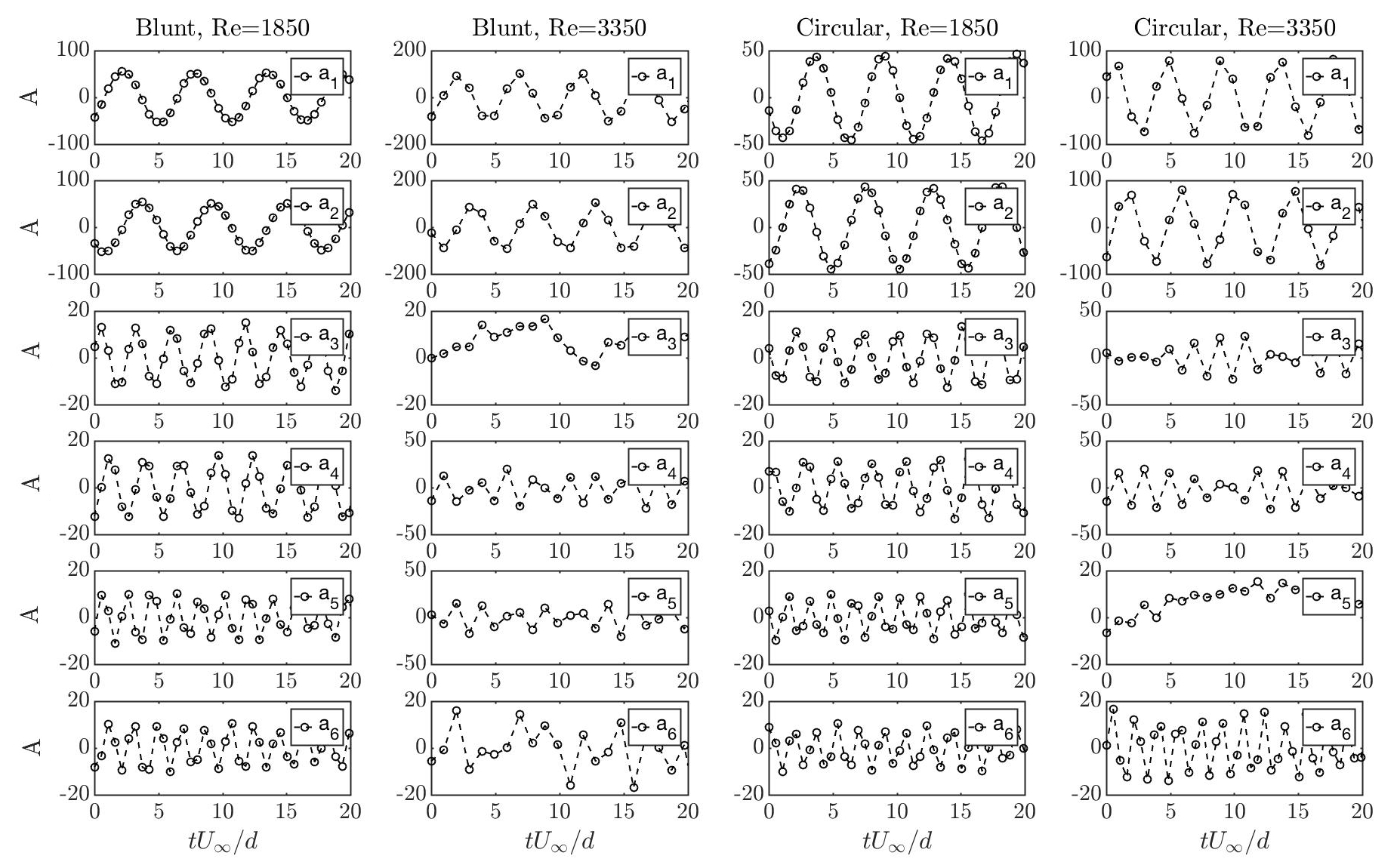}
	}
	\caption{The project POD coefficients, $a^{k}$, corresponding to the POD modes shown in Fig. \ref{fig:pod2}.
	}
	\label{fig:pod22}
\end{figure} 

  \begin{figure}
	\unitlength=95.0mm
	\centerline{
		\includegraphics[width=1.85\unitlength,height=1.15\unitlength]{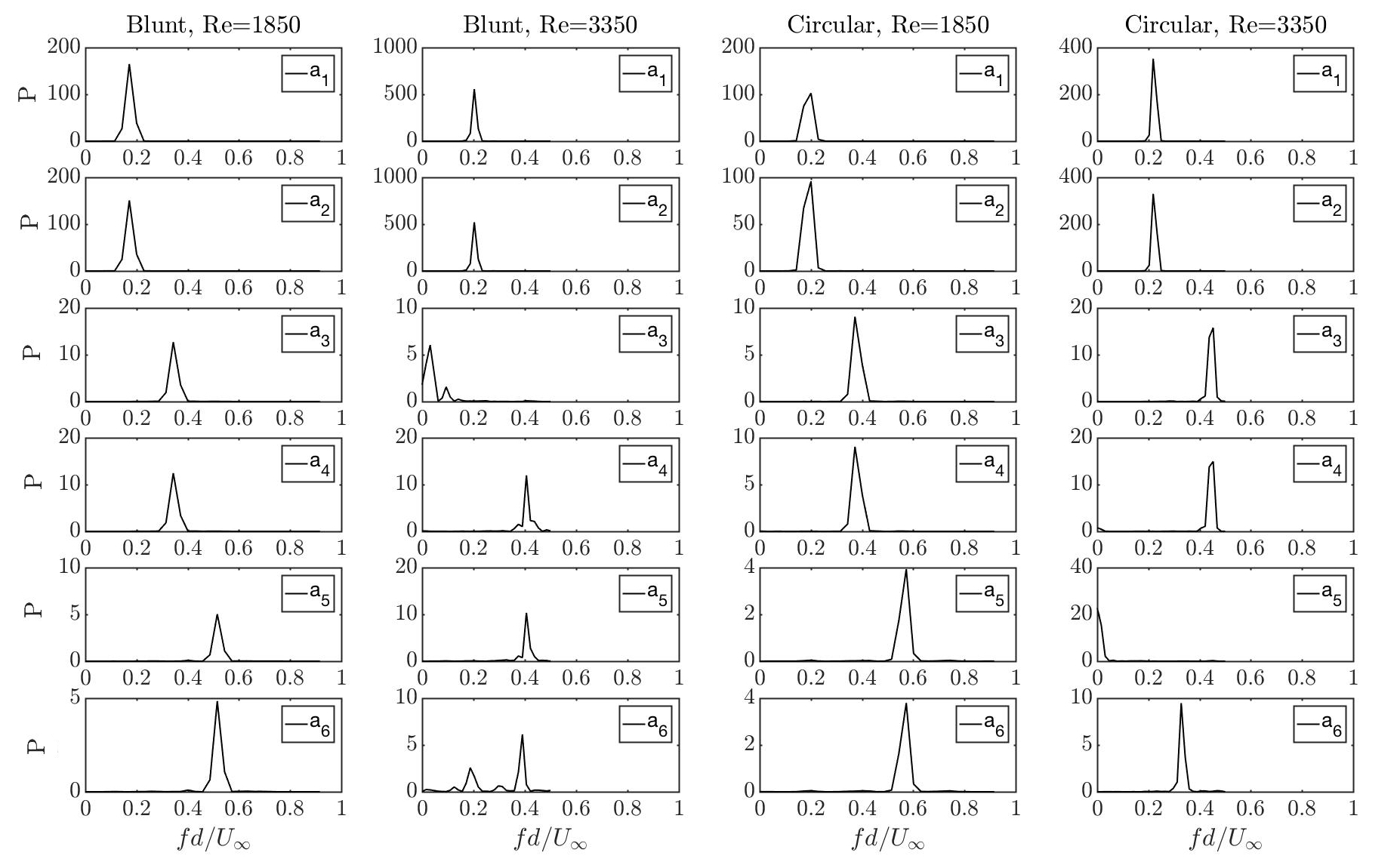}
	}
	\caption{Spectra of the POD coefficients, $a^{k}$, shown in Fig. \ref{fig:pod22}. $\textrm{P}$ indicates power in arbitrary scales.}
	\label{fig:pod222}
\end{figure}

% \begin{figure}
% \unitlength=70.0mm
%\centerline{
%\includegraphics[width=1.967\unitlength,height=1.112\unitlength]{Figures/POD_mode_coefficients_PRF.jpg}
%}
% \caption{Comparison of the simulated coefficients (solid line), $a_{1}$ and $a_{2}$, with their experimental counter parts (red filled circles), obtained from direct projection, for the different cases considered in the present study.
%}
% \label{fig:pod3}
% \end{figure} 

 \begin{table}
 	\caption{Relative and cumulative energy for the first 10 POD modes for four different cases.}
 	\begin{tabular}{|c|c|c|c|c|c|c|c|c|}
 		\tableline 
 	Mode&	Blunt & Blunt & Blunt & Blunt & Circular & Circular & Circular & Circular\\
 	&	$Re=1850$  & $Re=1850$ & $Re=3350$ & $Re=3350$ & $Re=1850$ & $Re=1850$ & $Re=3350$ & $Re=3350$\\
 		\tableline
 Number	&	Relative & Cumulative &  Relative & Cumulative & Relative & Cumulative & Relative & Cumulative\\
 	&	Energy (\%) & Energy (\%) &  Energy (\%) & Energy (\%) & Energy (\%) & Energy (\%) & Energy (\%) & Energy (\%)\\
 		\tableline
 	1 &	42.6062 & 42.6062 & 37.0986 & 37.0986 & 41.4926 & 41.4926 & 38.0795 & 38.0795\\
 		\tableline
 	2 &	38.6873 &	81.2934	& 34.4181 & 71.5167	& 38.0011&	79.4937&	36.1282& 74.2077\\
 		\tableline
    3 &	2.6852 & 83.9786	&1.8938&	73.4105&	2.1983&	81.692&	2.2511&	76.4589  \\
 		\tableline
 	4 &	2.6349 & 86.6135 & 1.6469&	75.0573 & 2.1657	&83.8577 &	2.138 &	78.5968 \\
 		\tableline
 	5 &	1.352 &	87.9655&	1.6044&	76.6617&	1.1788&	85.0365&	1.4845&	80.0813 \\
 		\tableline
 6 &		1.2912& 89.2567&	1.0747&	77.7364&	1.1075&	86.1441&	1.0535&	81.1348 \\
 		\tableline
 	7 &	0.8246& 90.0813& 	1.0381& 	78.7745& 	0.6634& 	86.8074& 	1.0034& 	82.1382 \\
 		\tableline
 8 &		0.5391&	90.6204&	0.9213&	79.6958&	0.6358&	87.4433&	0.7792&	82.9174 \\
 		\tableline
 9 &		0.4842&	91.1046&	0.9029&	80.5986&	0.5541&	87.9974&	0.7426&	83.66\\
 		\tableline
 10 &		0.4131&	91.5177&	0.7123&	81.311&	0.5075&	88.5048&	0.5091&	84.1691\\
 		\tableline
 	\end{tabular}
 	\label{tab:eigval}
 \end{table}

To quantify the interaction between the mean flow harmonic with the higher harmonics, which is the crux of the theoretical criterion derived in \cite{sipp2007global}, we look at the reduced order dimensionality of the flat plate wake. To this end, we compute the energy contribution of the different POD modes. As outlined in the previous section, the eigenvalue corresponding to each POD mode represents the proportional contribution to the total energy content. Fig. \ref{fig:pod1} describes the relative energy distribution of the different modes, for the different cases investigated in the present study. The numerical values of the relative and the cumulative energy of the first ten POD modes are also listed in table~\ref{tab:eigval}.  One may notice that the first two POD modes are the dominant ones. To be specific, the first two modes carry about $81.3\%, 71.5\%, 79.5\%$ and $74.2\%$ of the total energy, for blunt $Re=1850$, blunt $Re=3350$, circular $Re=1850$, circular $Re=3350$ cases, respectively. Table~\ref{tab:eigval} also shows that the first ten POD modes carry more than $80\%$ energy for all the cases. It also indicates that some eigenvalues come as pairs, for example, first two eigenvalues for all the cases; the corresponding modes are known as degenerate eigenmodes, which originate due to some symmetry in the flow~\cite{Dean91, Aubr92, remp94, Mand10}, as further discussed below. Similar observations have also been made in the case of cylinder wakes \cite{noac03,siegel_seidel_fagley_luchtenburg_cohen_mclaughlin_2008}.\\

The POD modes or the eigenfunctions corresponding to the first six eigenvalues are shown in Fig. \ref{fig:pod2}. The corresponding time coefficients obtained from direct projections (equation~\ref{eqn:coeff}) and their frequency spectra are shown in Fig. \ref{fig:pod22} and \ref{fig:pod222}, respectively. The first two near degenerate POD modes are nearly the same with a slight spatial and temporal shift as seen in Fig. \ref{fig:pod2} and \ref{fig:pod22}, respectively. This sort of shift associated with a pair of POD modes is attributed to some sort of symmetry/traveling disturbance in the flow. Here, this is due to the vortex shedding at the trailing edge of the flat plate, as the frequency of these disturbances in Fig. \ref{fig:pod222} (first row) matches with the shedding frequency in Fig. \ref{fig:freqexp}. Similarly, second and third pairs of modes corresponding to the second and third pairs of eigenvalues, for the blunt $Re=1850$ and the circular $Re=1850$ cases, show the first and the second harmonics associated with the vortex shedding. In contrast, one solitary mode (mode 3 and mode 5) which does not make any pair is seen to appear for the blunt and circular cases at $Re=3350$, respectively. This mode is similar to the shift mode as discussed in~\cite{noac03} for a circular cylinder. The spatial structure of this mode is shown in Fig.~\ref{fig:pod2} (second column 3rd mode, and fourth column fifth mode) and the corresponding time series and frequency of this mode in Figs.~\ref{fig:pod22} and~\ref{fig:pod222}, for the blunt and the circular cases at $Re=3350$. These observations indicate that this mode is associated with a slowly moving disturbance. These results show that the trailing edge geometry has little influence on the vortex shedding  characteristics, whereas the Reynolds number does play a significant role. Further, the present POD analysis reveals that the flat plate wake at this moderate Reynolds number is low dimensional in nature as the first ten POD modes can capture more than $80\%$ of the flow energy.\\

% It shows that the structure of the first two harmonics are very similar, with a shift in the streamwise direction for all the cases. Indeed, these two modes contribute the most to the total energy. The fact that their spatial distribution almost remains unaltered in all the cases investigated in the present study, shows that they are not significantly impacted by changes to the trailing edge geometry, and the Reynolds number. In addition, we observe that the POD modes for our cases appear in pairs Figs. (\ref{fig:pod1},\ref{fig:pod2}). The pairing of modes produced from the POD method is attributed to the coherent structures present in the flow which travel downstream  \cite{remp94}. This also suggests that the POD method for the flat plate wake would produce degenerate modes confirming their ability  to capture the coherent structures present in the flow.

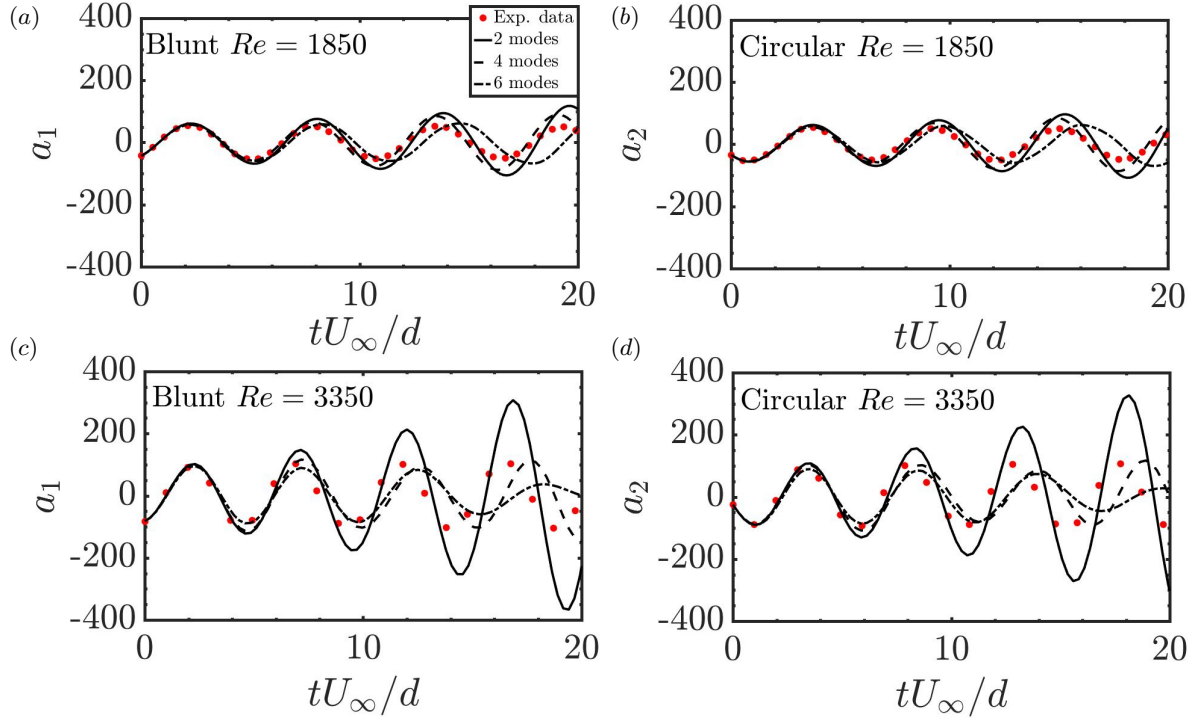
\begin{figure}
	\unitlength=80.0mm
	\centerline{
		\includegraphics[width=1.94\unitlength,height=1.2\unitlength]{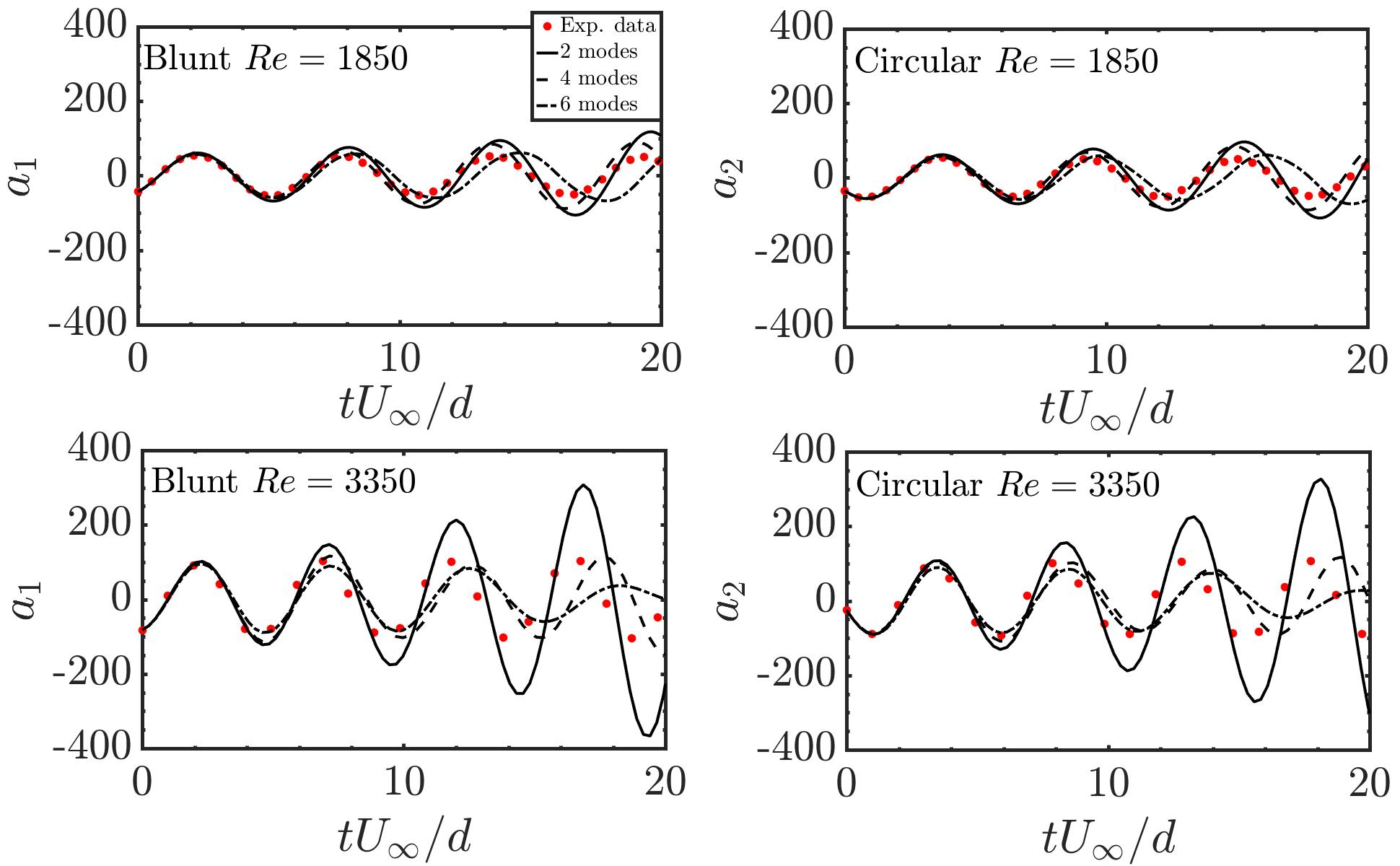}
	}
	\begin{picture}(0,0)
		\put(-1.0,1.2){$(a)$}
		\put(0.0,1.2){$(b)$}
		\put(-1.0,0.65){$(c)$}
		\put(0.0,0.65){$(d)$}
	\end{picture}
	\caption{Comparison of the simulated coefficients (solid line), $a_{1}$ and $a_{2}$, obtained using different mode models (i.e. 2, 4, 6 modes models), with their experimental counter parts (red filled circles), calculated from direct projection, for the different cases considered in the present study.
	}
	\label{fig:pod5}
\end{figure}

\begin{figure}
	\unitlength=80.0mm
	\centerline{
		\includegraphics[width=1.98\unitlength,height=1.21\unitlength]{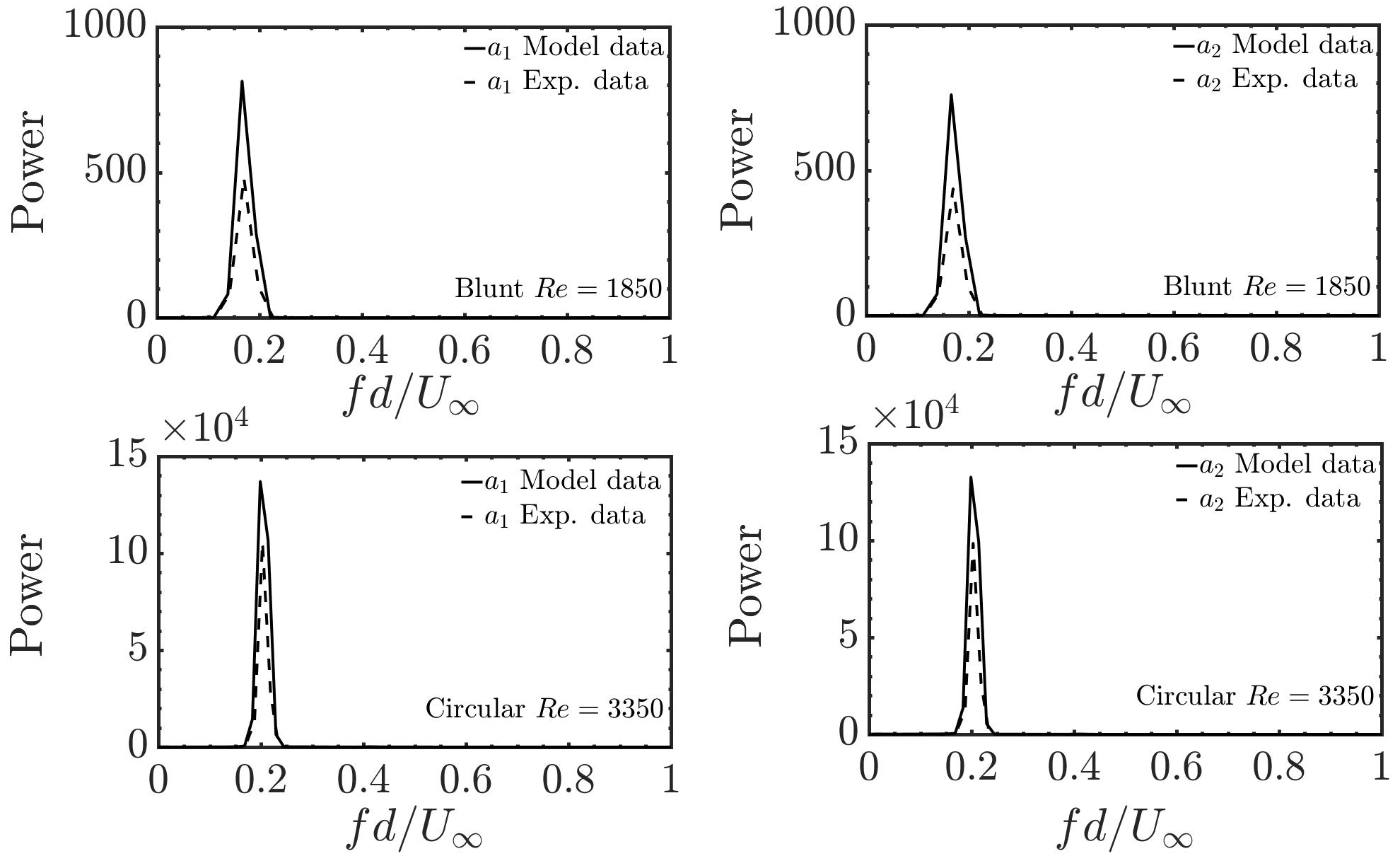}
	}
	\begin{picture}(0,0)
		\put(-1.0,1.2){$(a)$}
		\put(0.0,1.2){$(b)$}
		\put(-1.0,0.65){$(c)$}
		\put(0.0,0.65){$(d)$}
	\end{picture}
	\caption{Comparison of the frequency spectra of the time coefficients, $a_{1}$ and $a_{2}$ obtained from the simulation (solid line), and from the direct projection (dashed line). They are in excellent agreement in all the four different cases investigated.
	}
	\label{fig:pod4}
\end{figure} 

Since the vortex shedding dynamics is low dimensional, it is interesting to check whether a reduced order model/low dimensional model can describe the vortex shedding characteristics. Considering just two POD modes, the equation~\ref{eqn:dyn} was numerically solved in Matlab. The numerical solutions for $a_{1}$ and $a_{2}$ are compared with their experimental counter parts obtained using direct projection, as shown in Fig. \ref{fig:pod5}, for all the cases. It can be seen that the coefficients of the first two modes closely follow the experimental data. The simulated frequencies associated with the first two POD modes, and the experimentally measured global shedding frequency from the instantaneous velocity fields are shown in Fig. \ref{fig:pod4}. One may notice an excellent agreement between the numerical solution and the experimental data, at least for one shedding cycle. In fact, inclusion of more number of modes while solving the equation~\ref{eqn:dyn}, does not improve the solution significantly, as seen in Fig. \ref{fig:pod5}. Although the amplitudes are found to be comparable with the experimental data for the four and six mode models, there is a significant deviation of phase as the solution progresses in time, which is better seen for higher $Re$ cases considered here. \\

These findings reaffirm the fact that there is a strong interaction of the mean flow (zeroth harmonic) with the first harmonic of the flow unsteadiness as compared to the second and third harmonics of the flow unsteadiness. This is due to the fact the simulated results were obtained from the numerical solution of Eq.~\ref{eqn:dyn} using the mean velocity and the first two POD modes. This clearly signifies that the interaction strength between the zeroth and the first harmonic is the dominant one. This is consistent with the theoretical finding of \cite{sipp2007global}, and in turn explains why the mean flow velocity fields can give an accurate estimate of the global unsteadiness, thereby exerpimentally supporting \cite{sipp2007global}. It is pertinent to point out that the theoretical criterion outlined in \cite{sipp2007global} has been experimentally shown only for the case of a circular cylinder wake in \cite{khor2008global}.\\
\\

\section{Discussion and concluding remarks}
The global shedding frequency of the flat plate wake is well-predicted from a local spatiotemporal stability analysis based on the mean flow velocity fields. The mean flow velocity fields were obtained using a time resolved-particle image velocimetry (TR-PIV) technique in a low-speed wind tunnel. 
Two values of the Reynolds number, $Re$, based on the freestream velocity $U_{\infty}$ and the flat plate thickness $D$ (which is fixed throughout the study), were considered: $1850$ and $3350$. A blunt, and a circular profile is used for the trailing edge geometry of the flat plate. Though the time-mean velocity profiles at different streamwise locations can be obtained using the hotwire anemometry technique (as in \cite{khor2008global}), the present investigation uses a non-intrusive PIV measurement technique. This measurement technique is best suited for separating flows like the present one, and it is widely used for simultaneous whole field measurements as compared to the hotwire anemometry technique. 
The local stability analysis assumes that the flows are nearly parallel in the regions where the stability analysis is performed. Despite this, as observed in earlier studies, notably the vortex shedding from interacting boundary layers at a blunt trailing edge of a rectangular forebody in \cite{Hammond1}, the selected frequency is surprisingly well estimated with a deviation within $1\%$. In addition, the frequency at the end of the absolutely unstable region is seen to be selected by the global wake in the present study, which is the hydrodynamic resonance criterion advanced by \cite{koch1985local}. Similar observations have been detailed for a cylinder wake based on the mean flow \cite{Pier2,khor2008global}. 

As discussed in \cite{Pier2}, the linear global frequency corresponding to the streamwise locations where the flow changes the nature of instability from convective to absolute, cannot account for the real global frequency selected by the wake. In addition, the saddle-point criterion gave the best estimate for the shedding frequency of the fully developed cylinder wake when the stability analysis was applied on the time-averaged mean flow. This has been theoretically \cite{sipp2007global}, numerically \cite{barkley2006linear,mittal2008global}, and experimentally \cite{khor2008global} supported. The same characteristics have been observed in the present study based on time-averaged mean flow fields for a flat plate wake, with the frequencies obtained from a local analysis providing excellent predictions for the global shedding frequency. Interestingly, as observed in other wake flow studies, the local absolute frequency at the stagnation point was found to give an excellent estimate of the global shedding frequency. We would like to emphasize that to the best of our knowledge, this is the first experimental study to quantitatively investigate the global shedding frequency of a flat plate wake, and compare the different selection criteria available in literature using mean flow velocity fields.

To further support these observations, we have performed a POD analysis using the velocity fluctuations. It has been shown that the first two POD modes (which occur in pairs) account for over $70\%$ to the total energy of the flow. The frequency of the wake predicted using a low dimensional model with these two modes are in excellent agreement with the experimentally measured values. The physical mechanism behind the two theoretical conditions presented in \cite{sipp2007global} is related to the fact the frequency of the nonlinearly saturated limit cycle of the unsteady wake depends on the interaction of the mean flow with the higher harmonics. The fact that the first two POD modes dominate the total energy supports the criterion outlined in \cite{sipp2007global}. The findings from the present study should motivate further studies and possible control applications where the accurate estimation of the global shedding frequency of flat plate wakes are highly desired.

\section*{ACKNOWLEDGEMENTS}

D. D., I. K., and A. C. M. thankfully acknowledge the financial support provided by IIT Kanpur for the TR-PIV system. Dr. Balamurugan and Mr. Eswar Sunder are also thankfully acknowledged for their assistance with the data analyses and the TR-PIV measurements. They also thank Mr. Abhinath Kr. Yadav and Mr. Dorilal for their help with the experimental setup in the wind tunnel. S. S. G. acknowledges the support from the French National Research Agency (LABEX CEMPI, Grant No. ANR-11- LABX-0007) as well as the French Ministry of Higher Education and Research, Hauts de France council and European Regional Development Fund (ERDF) through the Contrat de Projets Etat-Region (CPER Photonics for Society P4S). 

\section*{REFERENCES}
 
%\section*{Appendix}
%\bibliographystyle{plain}
\bibliographystyle{unsrtnat}
\bibliography{Biblio_FPW_POD}

\begin{thebibliography}{79}
\providecommand{\natexlab}[1]{#1}
\providecommand{\url}[1]{\texttt{#1}}
\expandafter\ifx\csname urlstyle\endcsname\relax
  \providecommand{\doi}[1]{doi: #1}\else
  \providecommand{\doi}{doi: \begingroup \urlstyle{rm}\Url}\fi

\bibitem[Sipp and Lebedev(2007)]{sipp2007global}
D.~Sipp and A.~Lebedev.
\newblock Global stability of base and mean flows: a general approach and its
  applications to cylinder and open cavity flows.
\newblock \emph{J. Fluid Mech.}, 593:\penalty0 333--358, 2007.

\bibitem[Strouhal(1878)]{Strouhal1}
V.~Strouhal.
\newblock Uber eube besondere art der tonerregung.
\newblock \emph{Ann. d. Phys. u. Chem. N. F. (Leipzig)}, 5(10):\penalty0
  216--251, 1878.

\bibitem[Roshko(1955)]{Roshko1}
A.~Roshko.
\newblock On the wake and drag of bluff bodies.
\newblock \emph{J. Aeronaut. Sci.}, 22:\penalty0 124--132, 1955.

\bibitem[Bauer(1961)]{bauer1961vortex}
Andrew~B Bauer.
\newblock Vortex shedding from thin flat plates parallel to the free stream.
\newblock \emph{Journal of the Aerospace Sciences}, 28\penalty0 (4):\penalty0
  340--341, 1961.

\bibitem[Bloor(1964)]{Bloor1}
S.~Bloor.
\newblock The transition to turbulence in the wake of a circular cylinder.
\newblock \emph{J. Fluid Mech.}, 19:\penalty0 290--304, 1964.

\bibitem[Huerre and Redekopp(1990)]{Huerre1}
P.~Huerre and L.~Redekopp.
\newblock Local and global instabilities in spatially developing flows.
\newblock \emph{Annu. Rev. Fluid Mech.}, 22:\penalty0 473--537, 1990.

\bibitem[Williamson(1996)]{Will2}
C.~H.~K. Williamson.
\newblock Vortex dynamics in the cylinder wake.
\newblock \emph{Annu. Rev. Fluid Mech.}, 28:\penalty0 477--539, 1996.

\bibitem[Chomaz et~al.(1991)Chomaz, Huerre, and Redekopp]{Chomaz1}
J.~{-M}. Chomaz, P.~Huerre, and L.~Redekopp.
\newblock A frequency selection criterion in spatially developing flows.
\newblock \emph{Stud. Appl. Maths}, 84:\penalty0 119--144, 1991.

\bibitem[Williamson(1988)]{Will1}
C.~H.~K. Williamson.
\newblock Detuning a universal and continuous {S}trouhal--{R}eynolds number
  relationship for the laminar vortex shedding of a circular cylinder.
\newblock \emph{Phys. Fluids}, 31:\penalty0 2742--2744, 1988.

\bibitem[Monkewitz(1988)]{Monk1}
P.~Monkewitz.
\newblock The absolute and convective nature of instability in two-dimensional
  wakes at low reynolds numbers.
\newblock \emph{Phys. Fluids}, 31:\penalty0 999--1006, 1988.

\bibitem[Monkewitz et~al.(1993)Monkewitz, Huerre, and Chomaz]{Monk2}
P.~Monkewitz, P.~Huerre, and J.~{-M}. Chomaz.
\newblock Global linear stability analysis of weakly non-parallel shear flows.
\newblock \emph{J. Fluid Mech.}, 251:\penalty0 1--20, 1993.

\bibitem[Pier and Huerre(2001)]{Pier1}
B.~Pier and P.~Huerre.
\newblock Nonlinear self-sustained structures and fronts in spatially
  developing wake flows.
\newblock \emph{J. Fluid Mech.}, 435:\penalty0 145--174, 2001.

\bibitem[Pier(2002)]{Pier2}
B.~Pier.
\newblock On the frequency selection of finite-amplitude vortex shedding in the
  cylinder wake.
\newblock \emph{J. Fluid Mech.}, 458:\penalty0 407--417, 2002.

\bibitem[Hammond and Redekopp(1997)]{Hammond1}
D.~Hammond and L.~Redekopp.
\newblock Global dynamics of symmetric and asymmetric wakes.
\newblock \emph{J. Fluid Mech.}, 231:\penalty0 231--260, 1997.

\bibitem[Barkley(2006)]{barkley2006linear}
D.~Barkley.
\newblock Linear analysis of the cylinder wake mean flow.
\newblock \emph{Europhys. Lett.}, 75\penalty0 (5):\penalty0 750, 2006.

\bibitem[Mittal(2008)]{mittal2008global}
S.~Mittal.
\newblock Global linear stability analysis of time-averaged flows.
\newblock \emph{Int. J. Numer. Methods Fluids}, 58\penalty0 (1):\penalty0
  111--118, 2008.

\bibitem[Hwang and Choi(2006)]{Hwang1}
Y.~Hwang and H.~Choi.
\newblock Control of absolute instability by basic-flow modification in a
  parallel wake at low {R}eynolds number.
\newblock \emph{J. Fluid Mech.}, 560:\penalty0 465--475, 2006.

\bibitem[Gianetti and Luchini(2007)]{Luchini1}
F.~Gianetti and P.~Luchini.
\newblock Structural sensitivity of the first instability of the cylinder wake.
\newblock \emph{J. Fluid Mech.}, 581:\penalty0 167--197, 2007.

\bibitem[Sato and Kuriki(1961)]{Sato61}
H.~Sato and K.~Kuriki.
\newblock The mechanism of transition in the wake of a thin flat plate placed
  parallel to a uniform flow.
\newblock \emph{J. Fluid Mech.}, 11\penalty0 (03):\penalty0 321--352, 1961.

\bibitem[Ko et~al.(1970)Ko, Kubota, and Lees]{koku70}
D.~R.-S. Ko, T.~Kubota, and L.~Lees.
\newblock Finite disturbance effect on the stability of a laminar
  incompressible wake behind a flat plate.
\newblock \emph{J. Fluid Mech.}, 40:\penalty0 315–341, 1970.

\bibitem[Nishioka and Miyagi(1978)]{nish78}
M.~Nishioka and T.~Miyagi.
\newblock Measurements of velocity distributions in the laminar wake of a flat
  plate.
\newblock \emph{J. Fluid Mech.}, 84:\penalty0 705--–715, 1978.

\bibitem[Ramaprian et~al.(1982)Ramaprian, Patel, and Sastry]{rama82}
B.~R. Ramaprian, V.~C. Patel, and M.~S. Sastry.
\newblock The symmetric turbulent wake of a flat plate.
\newblock \emph{AIAA journal}, 20\penalty0 (9):\penalty0 1228--1235, 1982.

\bibitem[Wygnanski et~al.(1986)Wygnanski, Champagne, and Marasli]{wygn86}
I.~Wygnanski, F.~Champagne, and B.~Marasli.
\newblock On the large-scale structures in two-dimensional, small-deficit,
  turbulent wakes.
\newblock \emph{J. Fluid Mech.}, 168:\penalty0 31--71, 1986.

\bibitem[Jovic and Ramaprian(1986)]{jovi86}
S.~Jovic and B.~R. Ramaprian.
\newblock Large-scale structure of the turbulent wake behind a flat plate.
\newblock Technical report, Iowa Institute of Hydraulic Research, 1986.

\bibitem[Bogucz and Walker(1988)]{bogu88}
E.~A. Bogucz and J.~D.~A. Walker.
\newblock The turbulent near wake at a sharp trailing edge.
\newblock \emph{J. Fluid Mech.}, 196:\penalty0 555--584, 1988.

\bibitem[Julien et~al.(2003{\natexlab{a}})Julien, Lasheras, and Chomaz]{juli03}
S.~Julien, J.~Lasheras, and J.~{-M}. Chomaz.
\newblock Three-dimensional instability and vorticity patterns in the wake of a
  flat plate.
\newblock \emph{J. Fluid Mech.}, 479:\penalty0 155–189, 2003{\natexlab{a}}.

\bibitem[Lasheras and E.~Meiburg(1990)]{lash90}
J.~C. Lasheras and E~E.~Meiburg.
\newblock Three-dimensional vorticity modes in the wake of a flat plate.
\newblock \emph{Phys. Fluids}, 2\penalty0 (3):\penalty0 371--380, 1990.

\bibitem[Narasimha and Prabhu(1972)]{nara72}
R.~Narasimha and A.~Prabhu.
\newblock Equilibrium and relaxation in turbulent wakes.
\newblock \emph{J. Fluid Mech.}, 54\penalty0 (01):\penalty0 1--17, 1972.

\bibitem[Carini et~al.(2017)Carini, Airiau, Debien, L{\'e}on, and
  Pralits]{cari17}
M.~Carini, C.~Airiau, A.~Debien, O.~L{\'e}on, and J.~O. Pralits.
\newblock Global stability and control of the confined turbulent flow past a
  thick flat plate.
\newblock \emph{Phys. Fluids}, 29\penalty0 (2):\penalty0 024102, 2017.

\bibitem[A and Walker(1988)]{bogucz1988turbulent}
E.~A.~Bogucz A and J.~D.~A. Walker.
\newblock The turbulent near wake at a sharp trailing edge.
\newblock \emph{J. Fluid Mech.}, 196:\penalty0 555--584, 1988.

\bibitem[Nakamura et~al.(1991)Nakamura, Ohya, and
  Tsuruta]{nakamura1991experiments}
Y.~Nakamura, Y.~Ohya, and H.~Tsuruta.
\newblock Experiments on vortex shedding from flat plates with square leading
  and trailing edges.
\newblock \emph{J. Fluid Mech.}, 222:\penalty0 437--447, 1991.

\bibitem[Rai(2016)]{rai16}
M.~M. Rai.
\newblock Flat plate wake velocity statistics obtained with circular and
  elliptic trailing edges.
\newblock Technical report, 2016.

\bibitem[Sieverding and Heinemann(1990)]{Siev90}
C.~H. Sieverding and H.~Heinemann.
\newblock The influence of boundary layer state on vortex shedding from flat
  plates and turbine cascades.
\newblock \emph{J. Turbomach.}, 112:\penalty0 181--187, 1990.

\bibitem[Taylor et~al.(2011)Taylor, Palombi, Gurka, and Kopp]{Tayl11}
Z.~J. Taylor, E.~Palombi, R.~Gurka, and G.~A. Kopp.
\newblock Features of the turbulent flow around symmetric elongated bluff
  bodies.
\newblock \emph{J. Fluids Struct.}, 27\penalty0 (2):\penalty0 250--265, 2011.

\bibitem[Boldman et~al.(1976)Boldman, Brinich, and Goldstein]{Bold76}
D.~R. Boldman, P.~F. Brinich, and M.~E. Goldstein.
\newblock Vortex shedding from a blunt trailing edge with equal and unequal
  external mean velocities.
\newblock \emph{J. Fluid Mech.}, 75\penalty0 (4):\penalty0 721--735, 1976.

\bibitem[Rowe et~al.(2001)Rowe, Fry, and Motallebi]{Rowe01}
A.~Rowe, A.~L.~A. Fry, and F.~Motallebi.
\newblock Influence of boundary-layer thickness on base pressure and vortex
  shedding frequency.
\newblock \emph{AIAA journal}, 39\penalty0 (4):\penalty0 754--756, 2001.

\bibitem[Durgesh et~al.(2013)Durgesh, Naughton, and Whitmore]{Durg13}
V.~Durgesh, J.~W. Naughton, and S.~A. Whitmore.
\newblock Experimental investigation of base-drag reduction via boundary-layer
  modification.
\newblock \emph{AIAA journal}, 2013.

\bibitem[Koch(1985)]{koch1985local}
W.~Koch.
\newblock Local instability characteristics and frequency determination of
  self-excited wake flows.
\newblock \emph{J. Sound Vib.}, 99\penalty0 (1):\penalty0 53--83, 1985.

\bibitem[Pierrehumbert(1984)]{pierrehumbert1984local}
R.~T. Pierrehumbert.
\newblock Local and global baroclinic instability of zonally varying flow.
\newblock \emph{J. Atmos. Sci.}, 41\penalty0 (14):\penalty0 2141--2162, 1984.

\bibitem[Monkewitz and Nguyen(1987)]{monkewitz1987absolute}
P.~A. Monkewitz and L.~N. Nguyen.
\newblock Absolute instability in the near-wake of two-dimensional bluff
  bodies.
\newblock \emph{J. Fluids Struct.}, 1\penalty0 (2):\penalty0 165--184, 1987.

\bibitem[Dee and Langer(1983)]{dee1983propagating}
G.~Dee and J.~S. Langer.
\newblock Propagating pattern selection.
\newblock \emph{Phys. Rev. Lett.}, 50\penalty0 (6):\penalty0 383, 1983.

\bibitem[Khor et~al.(2008)Khor, Sheridan, Thompson, and
  Hourigan]{khor2008global}
M.~Khor, J.~Sheridan, M.~C. Thompson, and K.~Hourigan.
\newblock Global frequency selection in the observed time-mean wakes of
  circular cylinders.
\newblock \emph{J. Fluid Mech.}, 601:\penalty0 425, 2008.

\bibitem[Ryan et~al.(2005)Ryan, Thompson, and Hourigan]{Ryan05}
K.~Ryan, M.C. Thompson, and K.~Hourigan.
\newblock Three-dimensional transition in the wake of bluff elongated
  cylinders.
\newblock \emph{J. Fluid Mech.}, 538:\penalty0 1--29, 2005.

\bibitem[Naghib-Lahouti et~al.(2012)Naghib-Lahouti, Doddipatla, and
  Hangan]{Nagh12}
A.~Naghib-Lahouti, L.~S. Doddipatla, and H.~Hangan.
\newblock Secondary wake instabilities of a blunt trailing edge profiled body
  as a basis for flow control.
\newblock \emph{Exp. Fluids}, 52\penalty0 (6):\penalty0 1547--1566, 2012.

\bibitem[Naghib-Lahouti et~al.(2014)Naghib-Lahouti, Lavoie, and Hangan]{Nagh14}
A.~Naghib-Lahouti, P.~Lavoie, and H.~Hangan.
\newblock Wake instabilities of a blunt trailing edge profiled body at
  intermediate reynolds numbers.
\newblock \emph{Exp. Fluids}, 55\penalty0 (7):\penalty0 1779, 2014.

\bibitem[Doddipatla(2010)]{Dodd10}
L.~S. Doddipatla.
\newblock \emph{Wake dynamics and passive flow control of a blunt trailing edge
  profiled body}.
\newblock PhD thesis, The University of Western Ontario, 2010.

\bibitem[Julien et~al.(2003{\natexlab{b}})Julien, Ortiz, and
  Chomaz]{julien_lasheras_chomaz_2003}
S.~Julien, S.~Ortiz, and J.-M. Chomaz.
\newblock Three-dimensional instability and vorticity patterns in the wake of a
  flat plate.
\newblock \emph{J. Fluid Mech.}, 479:\penalty0 155--189, 2003{\natexlab{b}}.

\bibitem[Julien et~al.(2004)Julien, Ortiz, and Chomaz]{julien2004secondary}
S.~Julien, S.~Ortiz, and J.~M. Chomaz.
\newblock Secondary instability mechanisms in the wake of a flat plate.
\newblock \emph{Eur. J. Mech. B Fluids}, 23\penalty0 (1):\penalty0 157--165,
  2004.

\bibitem[Leweke and Williamson(1998)]{leweke1998three}
T.~Leweke and C.~H.~K. Williamson.
\newblock Three-dimensional instabilities in wake transition.
\newblock \emph{Eur. J. Mech. B Fluids}, 17\penalty0 (4):\penalty0 571--586,
  1998.

\bibitem[Barkley and Henderson(1996)]{barkley1996three}
Dwight Barkley and Ronald~D Henderson.
\newblock Three-dimensional floquet stability analysis of the wake of a
  circular cylinder.
\newblock \emph{Journal of Fluid Mechanics}, 322:\penalty0 215--241, 1996.

\bibitem[Huerre and Monkewitz(2000)]{Huerre2}
P.~Huerre and P.~A. Monkewitz.
\newblock \emph{Open shear flow instabilities. In {P}erspectives in {F}luid
  {D}ynamucs: a {C}ollective {I}ntroduction to {C}urrent {R}esearch (ed. {G}.
  {K}. {B}atchelor and {H}. {K}. {M}offatt and {M}. {G}. {W}orster)}.
\newblock Cambridge {U}niversity {P}ress, 2000.

\bibitem[Aubry et~al.(1988)Aubry, Holmes, Lumley, and Stone]{aubr88}
N.~Aubry, P.~Holmes, J.~L. Lumley, and E.~Stone.
\newblock The dynamics of coherent structures in the wall region of a turbulent
  boundary layer.
\newblock \emph{J. Fluid Mech.}, 192:\penalty0 115--173, 1988.

\bibitem[Noack et~al.(2003)Noack, Afanasiev, Morzynski, Tadmor, and
  Thiele]{noac03}
B.~R. Noack, K.~Afanasiev, M.~Morzynski, G.~Tadmor, and F.~Thiele.
\newblock A hierarchy of low-dimensional models for the transient and
  post-transient cylinder wake.
\newblock \emph{J. Fluid Mech.}, 497:\penalty0 335--363, 2003.

\bibitem[Berkooz et~al.(1993)Berkooz, Holmes, and Lumley]{berk93}
G.~Berkooz, P.~Holmes, and J.~L. Lumley.
\newblock The proper orthogonal decomposition in the analysis of turbulent
  flows.
\newblock \emph{Annu. Rev. Fluid Mech.}, 25\penalty0 (1):\penalty0 539--575,
  1993.

\bibitem[Rempfer and Fasel(1994)]{remp94}
D.~Rempfer and H.~F. Fasel.
\newblock Evolution of three-dimensional coherent structures in a flat-plate
  boundary layer.
\newblock \emph{J. Fluid Mech.}, 260:\penalty0 351–375, 1994.

\bibitem[Siegel et~al.(2008)Siegel, Seidel, Fagley, Luchtenburg, Cohen, and
  Mclaughin]{siegel_seidel_fagley_luchtenburg_cohen_mclaughlin_2008}
S.~G. Siegel, J.~Seidel, C.~Fagley, D.M. Luchtenburg, K.~Cohen, and
  T.~Mclaughin.
\newblock Low-dimensional modelling of a transient cylinder wake using double
  proper orthogonal decomposition.
\newblock \emph{J. Fluid Mech.}, 610:\penalty0 1--42, 2008.

\bibitem[Galletti et~al.(2004)Galletti, Bruneau, Zannetti, and Iollo]{gall04}
B.~Galletti, C.~H. Bruneau, L.~Zannetti, and A.~Iollo.
\newblock Low-order modelling of laminar flow regimes past a confined square
  cylinder.
\newblock \emph{J. Fluid Mech.}, 503:\penalty0 161--170, 2004.

\bibitem[Luchtenburg et~al.(2009)Luchtenburg, Noack, and Schlegel]{luch09}
D.~M. Luchtenburg, B.~R. Noack, and M.~Schlegel.
\newblock An introduction to the pod galerkin method for fluid flows with
  analytical examples and matlab source codes.
\newblock \emph{Berlin Institute of Technology MB1, Muller-Breslau-Strabe}, 11,
  2009.

\bibitem[Kanshana et~al.(2018)Kanshana, Sunder, and Mandal]{Mandal1}
I.~Kanshana, E.~Sunder, and A.~C. Mandal.
\newblock Low dimensional modeling of flow behind a flat plate with blunt
  trailing edge.
\newblock \emph{7th International and 45th National Conference on Fluid
  Mechanics and Fluid Power (FMFP)}, 2018.

\bibitem[Rempfer and Fasel(1993)]{remp93}
D~Rempfer and H~Fasel.
\newblock The dynamics of coherent structures in a flat-plate boundary layer.
\newblock \emph{Applied Scientific Research}, 51\penalty0 (1-2):\penalty0
  73--77, 1993.

\bibitem[Rajaee et~al.(1994)Rajaee, Karlsson, and Sirovich]{raja94}
M.~Rajaee, S.~K.~F. Karlsson, and L.~Sirovich.
\newblock Low-dimensional description of free-shear-flow coherent structures
  and their dynamical behaviour.
\newblock \emph{J. Fluid Mech.}, 258:\penalty0 1--29, 1994.

\bibitem[Balamurugan and Mandal(2017)]{Bala17}
G.~Balamurugan and A.~C. Mandal.
\newblock Experiments on localized secondary instability in bypass boundary
  layer transition.
\newblock \emph{J. Fluid Mech.}, 817:\penalty0 217--263, 2017.

\bibitem[Lourenco and Krothapalli(2000)]{Lour00}
L.~M. Lourenco and A.~Krothapalli.
\newblock {TRUE} resolution piv: a mesh-free second order accurate algorithm.
\newblock In \emph{Proceedings of the International Conference in applications
  of lasers to fluid mechanics, Lisbon, Portugal}. 2000.

\bibitem[Mandal et~al.(2010)Mandal, Venkatakrishnan, and Dey]{Mand10}
A.~C. Mandal, L.~Venkatakrishnan, and J.~Dey.
\newblock A study on boundary-layer transition induced by free-stream
  turbulence.
\newblock \emph{J. Fluid Mech.}, 660:\penalty0 114--146, 2010.

\bibitem[Mandal and Dey(2011)]{Mand11}
A.~C. Mandal and J.~Dey.
\newblock An experimental study of boundary layer transition induced by a
  cylinder wake.
\newblock \emph{J.~Fluid Mech.}, 684:\penalty0 60--84, 2011.

\bibitem[Phani~Kumar et~al.(2015)Phani~Kumar, Mandal, and Dey]{Phan15}
P.~Phani~Kumar, A.~C. Mandal, and J.~Dey.
\newblock Effect of a mesh on boundary layer transitions induced by free-stream
  turbulence and an isolated roughness element.
\newblock \emph{J.~Fluid Mech.}, 772:\penalty0 445--477, 2015.

\bibitem[Chomaz(2005)]{chomaz2005global}
J.~{-M}. Chomaz.
\newblock Global instabilities in spatially developing flows: non-normality and
  nonlinearity.
\newblock \emph{Annu. Rev. Fluid Mech.}, 37:\penalty0 357--392, 2005.

\bibitem[White and Corfield(2006)]{White06}
Frank~M White and Isla Corfield.
\newblock \emph{Viscous fluid flow}, volume~3.
\newblock McGraw-Hill New York, 2006.

\bibitem[Trefethen(2000)]{Trefethen1}
L.~N. Trefethen.
\newblock \emph{Spectral {M}ethods in {MATLAB}}.
\newblock {SIAM}, Philadelphia, 2000.

\bibitem[Bers(1961)]{Bers1}
A.~Bers.
\newblock \emph{Space-time evolution of plasma instabilities -- absolute and
  convective}.
\newblock In {H}andbook of {P}lasma {P}hysics, 1961.

\bibitem[Briggs(1964)]{Briggs1}
R.~Briggs.
\newblock \emph{Election-{S}tream {I}nteraction with {P}lasmas}.
\newblock {M}. {I}. {T} Press, 1964.

\bibitem[Sirovich(1987)]{siro87}
L.~Sirovich.
\newblock Turbulence and the dynamics of coherent structures. part i: Coherent
  structures.
\newblock \emph{Q. Appl. Math.}, 45\penalty0 (3):\penalty0 561--571, 1987.

\bibitem[Pedersen and Meyer(2002)]{Pede02}
J.~Pedersen and K.~Meyer.
\newblock Pod analysis of flow structures in a scale model of a ventilated
  room.
\newblock \emph{Experiments in fluids}, 33\penalty0 (6):\penalty0 940--949,
  2002.

\bibitem[Kanshana et~al.()Kanshana, Sunder, and Mandal]{indr18}
Indra Kanshana, Eswar Sunder, and A.~C. Mandal.
\newblock Low dimensional modeling of flow behind a flat plate with blunt
  trailing edge.
\newblock In \emph{Proceedings of the 7th International and 45th National
  Conference on Fluid Mechanics and Fluid Power}.

\bibitem[Asai et~al.(2002)Asai, Minagawa, and Nishioka]{asai2002instability}
M.~Asai, M.~Minagawa, and M.~Nishioka.
\newblock The instability and breakdown of a near-wall low-speed streak.
\newblock \emph{J. Fluid Mech.}, 455:\penalty0 289--314, 2002.

\bibitem[Pier et~al.(1998)Pier, Huerre, Chomaz, and Couairon]{pier1998steep}
B.~Pier, P.~Huerre, J.-M. Chomaz, and A.~Couairon.
\newblock Steep nonlinear global modes in spatially developing media.
\newblock \emph{Phys. Fluids}, 10\penalty0 (10):\penalty0 2433--2435, 1998.

\bibitem[Rees(2009)]{Rees2009}
S.~J. Rees.
\newblock \emph{{Hydrodynamic instability of confined jets and wakes and
  implications for gas turbine fuel injectors}}.
\newblock PhD thesis, University of Cambridge, 2009.

\bibitem[Deane et~al.(1991)Deane, Kevrekidis, Karniadakis, and Orszag]{Dean91}
A.~E. Deane, I.~G. Kevrekidis, G.~E. Karniadakis, and S.~A. Orszag.
\newblock Low-dimensional models for complex geometry flows: Application to
  grooved channels and circular cylinders.
\newblock \emph{Phys. Fluids}, 3\penalty0 (10):\penalty0 2337--2354, 1991.

\bibitem[Aubry et~al.(1992)Aubry, Guyonnet, and Lima]{Aubr92}
N.~Aubry, R.~Guyonnet, and R.~Lima.
\newblock Spatio-temporal symmetries and bifurcations via bi-orthogonal
  decompositions.
\newblock \emph{J. Nonlin. Sci.}, 2:\penalty0 183--215, 1992.

\end{thebibliography}

\end{document}